\documentclass[12pt,preprint]{aastex}

\slugcomment{submitted in ApJ}

\shorttitle{X-ray Emission of V1432 Aql}
\shortauthors{V. R. Rana, K. P. Singh, and P. Barrett}

\begin{document}


\title{X-Ray Emission and Optical Polarization
of V1432 Aquilae: An Asynchronous Polar}


\author{V.R. Rana\altaffilmark{1}, K.P. Singh}
\affil{Department of Astronomy \& Astrophysics,
Tata Institute of Fundamental Research, Homi Bhabha Road, Colaba,
Mumbai 400~005, INDIA}

\email{vrana@tifr.res.in, singh@tifr.res.in}

\author{P.E. Barrett}
\affil{Space Telescope Science Institute, ESS/Science Software Branch,
Baltimore, MD 21218, U.S.A.}
\email{barrett@stsci.edu}

\and

\author{D.A.H. Buckley}
\affil{South African Astronomical Observatory, PO Box 9, Observatory
7935, Cape Town, South Africa}
\email{dibnob@saao.ac.za}

\altaffiltext{1}{Joint Astronomy Programme, Department of Physics,
Indian Institute of Science, Bangalore 560~012, INDIA}

\begin{abstract} A detailed analysis of X-ray data obtained with $ROSAT$,
$ASCA$, $XMM$-$Newton$ and the $Rossi$ $X$-$ray$ $Timing$ $Explorer$
($RXTE$) for the asynchronous polar V1432~Aquilae is presented.  An
analysis of Stokes polarimetry data obtained from the South African
Astronomical Observatory (SAAO) is also presented.  Power spectra from
long-baseline $ROSAT$ data show a spin period of 12150~s along with
several frequency components related to the source.  However, the
second harmonic of the spin period dominates the power spectrum in the
$XMM$-$Newton$ data.  For the optical circular polarization, the
dominant period corresponds to half the spin period (or its first
harmonic).  The $ROSAT$ data can be explained as due to accretion onto
two hot spots that are not anti-podal.  The variations seen in the
optical polarization and the $ASCA$ and $XMM$-$Newton$ X-ray data
suggest the presence of at least three accretion foot prints on the
surface of the white dwarf.  Two spectral models, a multi-temperature
plasma model and a photo-ionized plasma model, are used to understand
the spectral properties of V1432~Aql.  The data from the $RXTE$
Proportional Counter Array (PCA) with its extended high energy response
are used to constrain the white dwarf mass to 1.2$\pm$0.1 M$_{\odot}$
using a multi-temperature plasma model.  The data from the European
Photon Imaging Camera (EPIC) on-board $XMM$-$Newton$ are well fitted by
both models.  A strong soft X-ray excess ($<0.8$ keV) is well modeled
by a blackbody component having a temperature of 80--90 eV.  The plasma
emission lines seen at 6.7 and 7.0 keV are well fitted using the
multi-temperature plasma model.  However, the fluorescent line at 6.4
keV from cold Fe requires an additional Gaussian component.  The
multi-temperature plasma model requires two absorbers: one that covers
the source homogeneously and another partial absorber covering $\sim
65$\% of the source.  The photo-ionized plasma model, with a range of
column densities for the Fe ions, gives a slightly better overall fit
and fits all emission line features.  The intensity and spectral
modulations due to the rotation of the white dwarf at a period of
12150~s require varying absorber densities and a varying covering
fraction of the absorber for the multi-temperature plasma model.  The
presence of a strong blackbody component, a rotation period of 12150~s,
modulation of the Fe fluorescence line flux with 12150~s period, and a
very hard X-ray component suggest that V1432~Aql is an unusual polar
with X-ray spectral properties similar to that of a soft intermediate
polar.  \end{abstract}


\section{Introduction}

V1432~Aquilae (RX J1940.1-1025) is an AM~Herculis (polar) system in
which a white dwarf with a strong magnetic field accretes material
from the Roche lobe of a late-type dwarf companion star.  It was
discovered in the ROSAT PSPC observations of the Seyfert galaxy
NGC~6814 (Madejski et al. 1993).  Staubert et al. (1994) detected
a period of 12120$\pm$3~s in radial velocity of the H$_{\alpha}$
emission line using optical spectroscopy.  The optical photometric
and X-ray light curves of V1432~Aql show an eclipse-like structure
near orbital phase zero that provided a more accurate orbital period
determination.  Watson et al. (1995) reported an orbital period of
12116.3$\pm$0.4~s using optical photometry and 12116.3$\pm$0.01~s
from combined optical and X-ray dip timings.  Further radial velocity
measurements by Patterson et al. (1995) and Friedrich et al. (1996)
confirmed this period.  Watson et al. (1995) argue that the eclipse-like
structure is due to absorption in the accretion stream (see also
Schmidt \& Stockman 2001).  Whereas, Patterson et al. (1995) suggest
that it is an eclipse by the secondary star.  Mukai et al. (2003a) and
Singh \& Rana (2003) (hereafter Paper 1) argue that this dip is likely
a true eclipse by the secondary star based on the absence of any
residual X-ray flux at minimum light.

The orbital period of V1432~Aql is well established from the well
defined eclipses.  The white dwarf spin period (P$_{spin}$) is
seen in the optical photometry (Patterson et al. 1995) as well as in
the X-ray data (Friedrich et al. 1996).  Geckeler \& Staubert
(1997) estimate the white dwarf spin period to be 12150~s from the
changing nature of the accretion region relative to the magnetic pole
in X-ray ($ROSAT$) and optical photometry.  Recently, Staubert et
al. (2003) have provided an accurate ephemeris for the spin period
using timings of spin minima obtained from the optical photometric and
the X-ray data and also detected a secular decrease of 1.013 $\times
10^{-8}$ s/s in the spin period.  They estimate a spin
period of 12150.328$\pm$0.045~s (for epoch 1995.2) using combined
optical and X-ray data.  Mukai et al. (2003a) report an analysis of
extensive optical photometry and derive an ephemeris of the spin
period (12150.432~s) and confirm the presence of its time derivative
\.P$_{spin}$=-9 $\times 10^{-9}$~s/s.  Based on these characteristics,
V1432~Aql is generally believed to be an asynchronous polar with the
unusual characteristic of the spin period being slightly longer than
the orbital period.  However, Mukai (1998) presented evidence for a
spin period of $\sim$4040~s using archival $ROSAT$ data and suggested
reclassification of V1432~Aql as an intermediate polar.  In paper 1,
Singh \& Rana supported this conclusion by reporting the strongest peak
of the power density spectrum at $\sim$4070~s from the analysis of a
continuous observation with $XMM$-$Newton$.  Additional
characteristics that further distinguish V1432~Aql from other
polars are the complex X-ray light-curves and the strong hard
X-ray emission (Singh et al. 2003), which are commonly associated with
the intermediate polars (IPs).

The accreting material in a polar or an IP forms a strong shock near
the surface of the white dwarf that heats the gas to a high
temperature ($\sim 10^8$ K).  The hot plasma in the post-shock region
gradually cools and settles on to the surface of the white dwarf,
emitting mainly in hard X-rays via thermal bremsstrahlung (Lamb \&
Masters 1979).
Therefore, the post-shock region can be considered to have a
continuous temperature distribution, so the use of a multi-temperature
plasma model to quantify the temperature distribution is necessary
(Done, Osborne \& Beardmore 1995).  The UV and soft X-ray components
(below 2 keV) mainly arise because of the reprocessing of hard X-ray
photons from the surface of the white dwarf.  The soft X-ray spectral
component can be modeled using blackbody emission that is
usually dominant in polars.  However, recently a few IPs
have been observed to show characteristics that are common to both IPs
and polars, and these are classified as soft IPs (Haberl \& Motch 1995).
These systems rotate asynchronously where the binary period
is considerably longer than the rotation period of the white
dwarf, and their X-ray light curves are strongly modulated at the spin
period.  In addition to a soft X-ray component they also show
polarized optical/IR emission due to electrons in the plasma
spiraling in a strong ($\sim 10^{7}$ B) magnetic field, properties
similar to that of polars.
Phase resolved X-ray spectroscopy of such systems
can be particularly useful to learn about the origin of the phase
dependent features and the complex absorption effects taking place in
the source, and the $XMM$-$Newton$ observatory with its wide energy
band and high sensitivity is an excellent instrument to study the
spectral nature and its phase-dependent behaviour in polars and IPs.
With its moderate energy resolution, $XMM$-$Newton$ can also provide
useful information about the Fe line complex.  Simultaneous
observations in the soft and hard X-ray energy bands provide important
information about the energy balance in these systems.

In this paper we present a detailed temporal and spectral analysis of
$ROSAT$, $ASCA$, $Rossi$ $X-ray$ $Timing$ $Explorer$ ($RXTE$) and
$XMM$-$Newton$ observations from 1992 to 2002.  We also present an analysis
of Stokes photo-polarimetry obtained at the $South$ $African$
$Astronomical$ $Observatory$ ($SAAO$) in 1994.  This paper is organized as
follows. In the next section we present the observations and analysis
techniques.  Section 3 contains results from the timing analysis including
the X-ray and the circular polarization power spectra and spin folded X-ray
light curves. Section 4 contains the results of the spectral analysis and a
description of spectral models that are used to characterize the source.
The results of the phase resolved spectroscopy are presented in \S 5. In \S
6, we discuss the timing and the spectral results, and compare them with
other polars and IPs.  We conclude and summarize the results in \S 7.

\section{Observations \& Data Analysis}

\subsection{$ROSAT$}

V1432~Aql was observed serendipitously with the $ROSAT$ (Tr\"{u}mper 1983)
Position Sensitive Proportional Counter (PSPC) in 1992--1993 during the
observations of a nearby Seyfert galaxy NGC~6814.  Data from these
observations were previously analyzed for temporal analysis by Staubert et
al. (1994), Friedrich et al. (1996), Mukai (1998), and Staubert et al.
(2003).  During the 1992 October observations, most of the data were
collected in last four days, hence for generating power spectra the data
from only the last four days have been used (i.e. 1992 October 27--30) (see
Mukai 1998).  The 1993 October observations were made at different offset
(17$\arcmin$) than the other two (37$\arcmin$). We have made no attempt to
account for the different offsets because the light curves are folded
separately and this will only affect the average intensity of the source
and not the overall profile of the intensity modulation and the intrinsic
variability of the source.  We extracted the background subtracted source
light curves in three energy bands of 0.1--0.5 keV (soft), 0.5--2 keV
(medium) and 0.1--2 keV (total) with a time resolution of 4~s, from a
circle with a radius of $2\arcmin$.5 centered on the peak position.  We
used the XSELECT program in FTOOLS ({\it see
http://heasarc.gsfc.nasa.gov/docs/software.html}) and performed a detailed
temporal analysis, exploiting the long baseline of the $ROSAT$
observations.

\subsection{$ASCA$}

The $ASCA$ observatory (Tanaka, Inoue, \& Holt 1994) observed
V1432~Aql serendipitously during 1997 October 27--29 while observing a
high redshift quasar, PKS 1937-101.
Here we have reanalyzed the Gas Imaging Spectrometer
(GIS2; Ohashi et al. 1996; Makishima et al. 1996) `PH' mode data with a
time resolution of 16~s using the standard screening criteria.
For further details the reader is referred to Mukai et al. (2003a).
The light curves were extracted in two energy bands of 0.7--2
keV (medium) and 2--10 keV (hard) to facilitate a direct
comparison with the light curves obtained from the $ROSAT$ and
the $XMM$ observations.

\subsection{$XMM$-$Newton$}

The $XMM$-$Newton$ observatory (Jansen et al. 2001) observed V1432~Aql {\it
continuously} on 2001 October 9 for 25455~s with the European Photon
Imaging Camera (EPIC) containing two MOS CCDs (Turner et al. 2001).  The
data from the EPIC PN camera were not available in the observing mode used
and the counts in the Reflection Grating Spectrometers (RGS1 and RGS2; den
Herder et al. 2001) were insufficient for a meaningful spectral analysis.
We adopted an energy range of 0.2--10 keV (a background flare affects the
data above 10 keV) for the light curve and time-averaged spectral analysis.
X-ray spectra and light curves were extracted for MOS1 and MOS2.  We use
the $rmfgen$ and $arfgen$ SAS (Science Analysis System ver. 5.4.1) tasks to
generate the appropriate response files for both MOS cameras.  Individual
spectra were binned to a minimum of 20 counts per bin to facilitate the use
of $\chi^2$ minimization during the spectral fitting.  X-ray light curves
were extracted in 4 energy bands; soft (0.2--0.5 keV), medium (0.5--2.0
keV), hard (2--10 keV), and total (0.2--10 keV) such that the soft and
medium bands correspond roughly to the $ROSAT$ soft and medium bands,
respectively.  $XMM$-$Newton$ light curves were presented in Figure 1 of
Paper 1.

\subsection{$RXTE$}

V1432~Aql was observed twice with the $RXTE$ Proportional Counter Array
(PCA; Jahoda et al. 1996) during 2002 July 14--15 (PI: P. Barrett).  The
observations were done at an offset of $\sim$15$\arcmin$ from the source to
minimize the contamination from the Seyfert galaxy NGC~6814 which is $\sim
37 \arcmin$ away.  At this offset the contamination from NGC~6814 is $\sim
10$\%.  The data were processed using the FTOOLS version 5.2 following the
standard procedure.  We analyzed the PCA standard 2 data with a time
resolution of 16~s.  For an improved signal-to-noise ratio, data in the
2--10 keV energy band from top xenon layer of all proportional counter
units (PCUs) were used to generate the source and the background light
curves and spectra.  The background light curve and spectrum were obtained
using the appropriate background model corresponding to the epoch of each
observations.  Only three of the five PCUs were ON during the two
observations.  However, data from PCU0 were rejected, because a propane
leak resulted in a large background flare affecting the top layer data from
PCU0.  Data from only 2 PCUs which are free from the background flare, were
analyzed.

\subsection{Stokes Polarimetry}

Stokes (simultaneous linear and circular) polarimetry data of V1432~Aql
were obtained using the University of Cape Town (UCT) polarimeter (Cropper
1985) on the 1 m telescope of the SAAO during 1994 July 5, 6, 8, and
10--13.  Most of the data were taken in white light (no filter), though
there were short periods of data taken with the B, V, Cousins R, Cousins I,
and an extended red (OG570) filters.  The integration time was 10~s for the
photometry and $\sim$5 minutes for the polarimetry.  The UBVRI filters were
calibrated before each run using standard stars and the waveplate offset
angles were calibrated using linear polarization standard stars.
Background exposures were taken at regular intervals to determine the net
flux and polarization.

\section{Timing Analysis}

The three $ROSAT$ observations during 1992--1993 with a mean separation of
about six months, provide sufficient frequency resolution to search for
fundamental periods, their harmonics and side-bands.  The frequency
resolution from the combined $ROSAT$ data is $8.13 \times 10^{-9}$ s$^{-1}$
and that from the $XMM$-$Newton$ data it is $9.83 \times 10^{-6}$ s$^{-1}$.
However, the $XMM$ data continuously span two complete binary orbits, while
those of $ROSAT$ data have gaps due to Earth occultation, SAA passage and
frequent switching of targets.  These gaps lead to complications in the
power spectrum as the true variations in the source are further modulated
by the irregular sampling defined by the window function of the data.  We
have used four methods to search for the periodicity present in the $ROSAT$
data: the discrete Fourier transform, the Lomb-Scargle periodogram (Horne
\& Baliunas 1986), the Bayesian statistical approach (Rana et al. 2004) and
the one-dimensional ``CLEAN'' algorithm (Roberts, Lehar, \& Dreher 1987) as
implemented in the PERIOD program (ver. 5.0-2) available in the STARLINK
package.  The Bayesian approach is mathematically more rigorous and can
provide optimum number of frequencies allowed by the data independent of
the shape of the periodic signal.  In the CLEAN algorithm, the discrete
Fourier transform first calculated from the light curve is followed by the
deconvolution of the ``window'' function from the data.  Here the power is
defined as the square of the half amplitude at the specified frequency.
This method has been demonstrated to work effectively for X-ray data from
various satellites (Norton et al. 1992a, 1992b; Rana et al. 2004).  We
applied the barycentric correction to all the data sets before carrying out
the power spectral analysis.  In addition, the data were detrended by
subtracting a mean count rate prior to generating power spectra using above
methods.

\subsection{Power Spectra} \subsubsection{$ROSAT$ PSPC} 
\begin{figure}
\epsscale{1.0}
\plotone{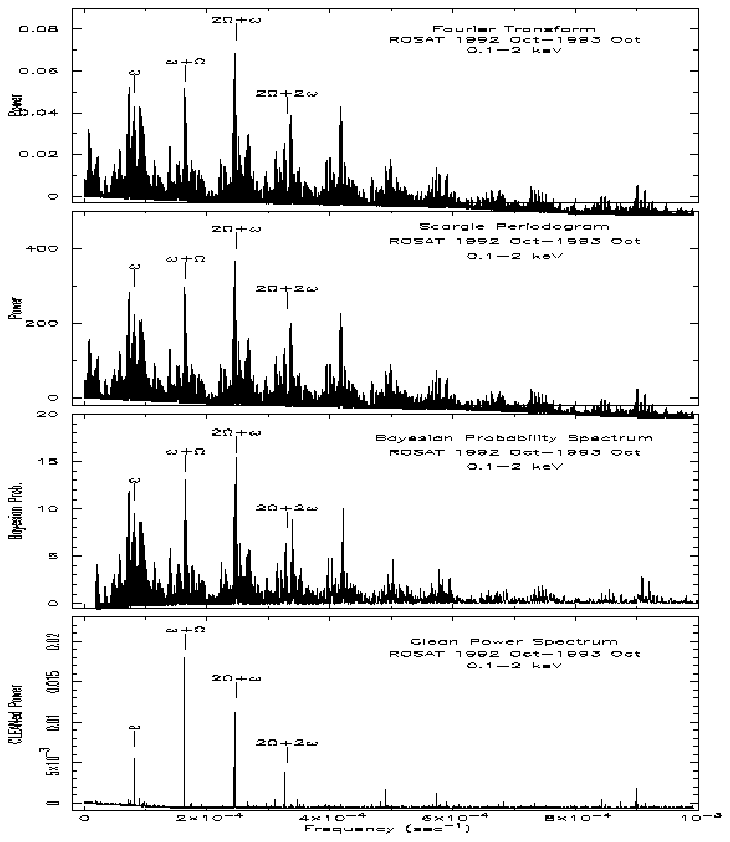}
\caption{Power density
spectra of V1432~Aql in the energy range 0.1--2 keV from combined
$ROSAT$ observations. First three panels show power spectra obtained
using discrete Fourier transform, Lomb-Scargle Periodogram, and
Bayesian method, respectively.  The bottom panel shows a CLEANed power
density spectrum.  Various identified frequency components are marked.
\label{rosatpds}} \end{figure}

The power spectra
obtained using above mentioned methods for the combined $ROSAT$ data in the
0.1--2 keV energy band, are shown plotted in the Figure \ref{rosatpds}.
All power spectra show four common significant peaks related to one of the
fundamental period and various side bands of the two fundamental periods of
the system, as marked in the figure.  The two fundamental frequencies of
the system are referred to as $\omega$, corresponding to the presumed spin
period and $\Omega$ corresponding to the orbital period. The power spectra
presented in first three panels of Fig. 1 show equally prominent peaks at
the two fundamental periods, suggesting that a combination of orbital and
spin modulations are at work in V1432~Aql. However, a significant peak is
seen at a frequency of 12150.72$\pm$1.14~s in CLEAN power spectrum, which
is consistent with the $\omega$ that is reported by Mukai et al. (2003a)
and Staubert et al. (2003).  No significant power is observed at $\Omega$,
which has a period of 12116.2824$\pm$0.0043~s (Mukai et al. 2003a),
suggesting that the data are modulated primarily on spin period.  The other
dominant peaks in the power spectra correspond to the $\Omega$+$\omega$,
2$\Omega$+$\omega$, and 2$\Omega$+2$\omega$ components.

In order to study the variations in the power density spectrum with energy,
we also calculated the power spectra for the soft and medium energy bands
(see \S 2.1).  The power spectra for the soft X-ray energy band show a
dominant peak at the 2$\Omega$+$\omega$ component.  The second dominant
peak, with almost equal power to that of first one, corresponds to the
2$\omega$ component, the first harmonic of the spin period.  Prominent peak
is also seen at $3\Omega+3\omega$ component.  The power spectra for the
medium energy X-ray band show a dominant peak at the 2$\omega$-$\Omega$
component.  The other dominant peaks correspond to frequency components
2$\Omega$+2$\omega$, 2$\omega$+$\Omega$, $\Omega$+$\omega$, and $\omega$ in
order of decreasing power.

\subsubsection{$XMM$-$Newton$ MOS}

Discrete Fourier transforms are performed on the X-ray light curves with
1~s time resolution.  The power spectra in the various energy bands of the
$XMM$-$Newton$ data were published in Paper 1.  However, the peaks in the
power spectra need to be reassigned based on the spin and orbital periods
being 12150.432~s and 12116.2824~s (Mukai et al. 2003a), respectively.  The
power spectrum obtained from averaged MOS1 and MOS2 data in the energy
range of 0.2--2 keV is shown plotted in Figure ~\ref{xmmpds}.  The energy
range is chosen so that this power spectrum can be directly compared with
that of $ROSAT$ total energy band.  However, the shorter baseline gives
broader peaks in the power spectrum and several frequency components are
indistinguishable. For example, the two fundamental frequencies $\omega$
and $\Omega$ have their power merged in to a single peak.  In addition,
several harmonics and side band frequencies of the fundamental periods are
merged together, they are 2$\omega$ and $\Omega$+$\omega$; 3$\omega$,
3$\Omega$, $\Omega$+2$\omega$ and 2$\Omega$+$\omega$; 4$\omega$, 4$\Omega$,
$\Omega$+3$\omega$ and 3$\Omega$+$\omega$; and 9$\omega$ and 9$\Omega$ as
shown in Figure 2.

\subsubsection{Circular Polarization in the Optical}

The polarization data are noisier as compared to the X-ray data and suffer
from poor and uneven sampling.  As a result the CLEAN procedure is not as
effective here as for X-ray data, and when used to generate the power
density spectrum from the polarization data, produces several strong peaks
at unexplained periods.  The power in the unidentified peaks dominates over
peaks at true system related periods.  Therefore, only a discrete Fourier
transform is presented (Figure~\ref{circpds}) for the circular polarization
data.  The most dominant peak in the power spectrum corresponds to a 1 day
positive alias of the 2$\omega$ component and is marked as $A^+_{2\omega}$.
Several peaks can be seen coincident with the fundamental frequency
$\omega$ and its various harmonics i.e. 2$\omega$, 3$\omega$, 4$\omega$ and
5$\omega$.  Power in the peak at the 5$\omega$ component is higher than at
3$\omega$ and 4$\omega$.  Several other unidentified peaks with significant
power, are also present in the power spectrum.

\subsection{Spin Folded X-ray Light curves}

To study variations in the X-ray intensity of the source as a function of
spin phase, we have folded the X-ray light curves on the spin period of the
system.  The X-ray light curves folded with the orbital period are already
studied in previous publications (see Mukai et al. 2003a and Paper 1).
When folding the X-ray data we have tried two spin ephemerides as given by
Mukai et al. (2003a) and Staubert et al. (2003), including the effect of
$\dot{P}$.  We find that the spin ephemeris as given by Mukai et al.
(2003a) better defines the spin minima for the X-ray data from recent years
(after 1997), and that of Staubert et al. (2003) is well suited for the
data from the earlier years (before 1997).  Since Mukai et al. (2003a)
obtained the spin ephemeris using the optical data spanning 7 years of base
line with clustering around 1998, that could be the reason for its better
fit to the spin minima for the data from the $ASCA$, $XMM$-$Newton$ and
$RXTE$.  On the other hand, Staubert et al. (2003) have used optical and
X-ray data taken during 1992--1997 to calculate the spin ephemeris, and
that could account for a better definition of the spin minima for the
$ROSAT$ data. With respect to a reference zero point defined using Mukai's
ephemeris, the spin folded features obtained using Staubert's ephemeris
show a varying phase shift from a negative value of -0.13 during 1992
October to a positive value of +0.1 during 2002 July. Spin folded light
curves in the three energy bands of soft, medium and hard X-rays are
presented in Figures~\ref{softlc}, \ref{medlc} and \ref{hardlc},
respectively.  Their main features are described below.

\subsubsection{Soft X-rays} Spin folded soft X-ray light curves for the
three $ROSAT$ observations in the 0.1--0.5 keV energy band (first three
panels) and for the $XMM$-$Newton$ observation in the 0.2--0.5 keV energy
band (bottom panel) are shown in Figure~\ref{softlc}.  The light curves
during 1992 October and 1993 March show similar profiles with a double hump
structure and a low intensity state during the 0.3--0.7 phase.  The
prominent humps span a phase range of 0.7--0.95 during 1992 October, and
0.7--0.9 during 1993 March.  The secondary humps (with smaller amplitude)
are of similar width as that of the primary but show a small phase shift
during the two observations.  The broad dip around the phase zero contains
two small peaks during 1992 October and is flat and relatively narrower
during 1993 March.  The light curve for the 1993 October observation (third
panel) is highly variable showing three narrow prominent peaks around
phases 0.2, 0.75, and 0.88 and a low intensity state during the
$\sim$0.25--0.7 phase range with a small flat bottomed dip around phase
0.37.  There is a broad dip between the phase range $\sim$0.95--1.12 that
is split into two small dips by an interpulse at phase 1.08.  For the
$XMM$-$Newton$ data, the profile based on ephemeris from Mukai et al.
(2003a) (solid line) shows a systematic shift of phase $\sim$0.1 with
respect to the profile based on ephemeris from Staubert et al. (2003)
(dash-dotted line).  The folded light curves show three peaks and three
dips as opposed to the double hump profile of the $ROSAT$ light curves of
1992 October and 1993 March, but similar to the $ROSAT$ 1993 October light
curves, albeit, with different widths and locations for the humps.  The
three minima are at phases 0.35, 0.65, and 1.05, respectively.  The width
of the minimum increases with the increasing spin phase.  The minimum at
0.65 is flat bottomed with a very sharp drop and a slow rise of the
intensity, whereas the other two minima have roughly the same falling and
rising times.  The third minimum at phase 1.05 is completely flat bottomed
with a constant intensity during 0.95--1.15.

\subsubsection{Medium Energy X-rays} The spin folded light curves in the
0.5--2 keV energy band for $ROSAT$, $XMM$-$Newton$, and in the 0.7--2 keV
energy band for the $ASCA$ observations are shown plotted in
Figure~\ref{medlc}.  The two different line styles represent folded
profiles for the same two ephemerides as used in Fig.~\ref{softlc}.  The
X-ray light curves in the first two $ROSAT$ observations show broadly
similar features with a double hump profile.  The width of the prominent
humps are identical but show a small phase shift during the 1992 October
and 1993 March observations, respectively, while the broader asymmetric
peaks cover phase ranges of $\sim$0.6--0.95 and $\sim$0.6--0.9.  The
intensity of the source is low but highly structured during the phase range
$\sim$0.3--0.6.  The spin profile during the third $ROSAT$ observation is
highly variable showing several peaks and dips unlike in the first two
observations.  The maximum of a prominent peak at phase 0.88 coincides with
the broad asymmetric peaks in the first two panels.  A second peak near
phase 0.2 resembles a peak during the 1992 October light curve but with a
relatively low amplitude. The low intensity phase of the source is narrower
and not as flat as that shown in the first two panels.  The broad dip
around phase zero is quite similar to that observed during 1992 October
with the presence of an interpulse at phase 1.08.  The $ASCA$ light curve
shows three humps -- one at phase 0.15 and two with relatively lower
amplitudes at phases 0.45 and 0.7, respectively.  The folded light curve
obtained from the $XMM$-$Newton$ data also shows three humps, with an
increasing amplitude with increasing spin phase.  The dip at phase 0.35 is
narrower and shallower than the dip at 0.65. During the broadest dip around
1.05 the intensity is not completely flat, instead shows a slow rise.  The
peak at phase $\sim$0.45 coincides with that observed in the $ASCA$ light
curve, whereas the other two peaks show a small phase shift.

Although overall profiles of the medium and soft energy band light curves
are roughly similar, the width and the amplitude of the peaks and dips are
different.  For the first two $ROSAT$ observations, the secondary peak in
the soft X-ray light curves becomes more prominent in the 0.5--2 keV band
while the widths remain unchanged; however the primary peak becomes broader
and asymmetric. On the other hand, during the 1993 March observations, both
the width of the peaks and their relative intensity are different in the
two bands.  Similarly, for the $XMM$-$Newton$ observations, the relative
amplitudes of the peaks in the medium energy band are different than in the
soft X-ray band but the widths of the peaks remain almost the same.  The
power spectra and light curves in different energy bands show that the
$ROSAT$ data are primarily modulated on the spin period, confirming the
claim by Staubert et al. (2003).  On the other hand, Mukai et al. (2003)
reported that the hard X-ray $BeppoSAX$ data are modulated on the orbital
period.

\subsubsection{Hard X-rays} Figure~\ref{hardlc} shows the folded profiles
in the 2--10 keV energy band for the $ASCA$, the $XMM$-$Newton$ and the
$RXTE$ observations, using the spin ephemeris as given by Mukai et al.
(2003a).  The $ASCA$ light curve shows three humps separated by three dips
with the humps located at similar phases as in the medium energy band. The
$XMM$-$Newton$ light curve, has a profile that is broadly similar to that
in the other two energy bands (Fig.~\ref{softlc} \& \ref{medlc}); the only
significant difference being in the shape of the minimum at phase 0.65.  In
hard X-rays, it is very sharp and narrow, suggesting a rapid uncovering of
the hard X-ray source, while the softer X-ray source remains covered.
Because of the insufficient exposure time, the $RXTE$ data could not cover
the entire spin cycle and there are gaps in the folded light curve. The
$RXTE$ light curve shows a broad peak during phase 0.6--0.75 followed by a
sharp drop ($\sim$17 times) in the intensity.  This low intensity phase
continues at least up to the 0.93 spin phase. Two more peaks are seen near
phases 0.12 and 0.22, the second of these coinciding with the peak observed
in the $XMM$-$Newton$ light curve.  Both the $XMM$-$Newton$ and the $RXTE$
hard X-ray profiles show two dips at spin phases $\sim$0.17 and $\sim$0.35,
respectively.

\subsection{Spin Folded Optical Polarization Curves}

Figure~\ref{optlc} represents the optical Stokes polarimetry data folded on
the spin period using the ephemeris as given by Staubert et al. (2003).  We
have combined the data from all nights to produced the average binned
polarization curves, since the individual unbinned polarization light
curves show almost the same trends.  The linear polarization shows a broad
peak covering a phase range of $\sim$0.45--0.80 that is suddenly cut by a
narrow dip around phase 0.6.  The position angle remains almost constant
over a phase interval of 0.0--0.4 with an average value of
$\sim$104$^{\circ}$.  It shows a broad asymmetric dip during phase 0.4--1.0
with a fast decline to a minimum value of 50$^{\circ}$ and a slow rise to a
maximum value of $\sim$140$^{\circ}$.  The circular polarization data shows
modulation over the spin period with two peaks -- a narrow peak around
phase 0.1 and a broad peak around phase 0.6.  Both positive and negative
values ranging from -1\% to 2\% are seen.

\section{X-ray Spectral Analysis}

The time-averaged X-ray spectra obtained from the $RXTE$ and the
$XMM$-$Newton$ observations are shown in Figures~\ref{rxtespec},
\ref{xmmsacg} \& \ref{xmmphoto}.  The overall shape of the spectra obtained
from the two $RXTE$ observations is quite similar except that the
normalized intensity is slightly lower during 2002 July 15.  The $RXTE$
spectra are very hard and extend to $\sim$30 keV.  They also show a broad
emission feature between 6 and 7 keV.  The $XMM$-$Newton$ spectra show the
presence of three emission lines at 6.4, 6.7, and 7.0 keV, with the first
line being the strongest.  Strong soft X-ray emission is also present in
the $XMM$-$Newton$ spectra.

Various plasma models have been used to describe the X-ray spectra of the
magnetic CVs (MCVs).  These include thermal bremsstrahlung, MEKAL (Mewe et
al. 1985; Liedahl et al. 1995), CEVMKL, and cooling flows as given in
XSPEC.  The MEKAL plasma code describes X-ray emission lines from several
elements and continuum expected from hot optically thin plasma assuming a
single temperature for the X-ray emitting plasma.  The CEVMKL is a
multi-temperature version of the MEKAL code with the emission measure
defined as a power law in temperature (Singh et al.  1996).  The models
include line emission from elements like He, C, N, O, Ne, Na, Mg, Al, Si,
S, Ar, Ca, Fe, and Ni.  None of these models is a true representation of
the physical conditions that exist in the X-ray emission regions of the
MCVs.  Spectral models, which can better describe the physical conditions
in MCVs, are the multi-temperature plasma model (model A) as described by
Cropper et al. (1999), and the photo-ionized plasma model (model B) as
described by Kinkhabwala et al. (2003).  Ramsay et al. (2004a) have used
model A to analyze the spectra of three polars that were observed by the
$XMM$-$Newton$ EPIC detectors (see the references therein for the use of
this model to other polar systems).  Similarly, Mukai et al. (2003b) have
used model B to describe the high resolution spectra of three IPs obtained
with the High Energy Transmission Grating (HETG) on-board the $Chandra$
observatory.  In this paper, we apply model A to the $RXTE$ spectra, and
models A and B to the $XMM$-$Newton$ MOS spectra.

\subsection{Multi-temperature Plasma Model}

The details about the formulation and the assumptions used in the
multi-temperature plasma model are described in Cropper et al. (1999).  The
model was obtained via private communication from G. Ramsay in 2003.  The
model predicts the temperature and the density profile of a magnetically
confined hot plasma in the post-shock region based on the calculations of
Aizu (1973).  It uses the `MEKAL' plasma code available in XSPEC to
describe the continuum and the line emission from a collisionally ionized
hot plasma.  Additionally, it takes into account the cyclotron cooling
effects that are important for MCVs.  The model includes the reflection of
hard X-rays from the white dwarf surface and the effects of the
gravitational potential on the structure of the post-shock region assuming
a stratified accretion column and a one-dimensional accretion flow.  The
model uses the mass-radius relationship of Nauenberg (1972) to estimate the
mass of the white dwarf.  It has a total of 8 free parameters.  They are
the ratio of time scales of the cyclotron cooling to the bremsstrahlung
cooling ($\epsilon_0$), the specific accretion rate ($\dot{m}$), the radius
of the accretion column ($r_c$), the mass of the white dwarf ($m_1$), the
Fe abundance relative to solar, the number of vertical grids in which the
shocked region is divided, the viewing angle to the reflecting site, and
the normalization.  To minimize the number of free parameters, some of the
system-dependent parameters are held constant.  In this analysis, the
viewing angle is fixed to $75^{\circ}$ and the post-shock region is divided
into 100 vertical layers.  Changing the value of viewing angle over a wide
range does not significantly affect the spectral fit.  By allowing the
remaining parameters to vary, the best fit was found when
$\epsilon_0\simeq$0.001, which is suitable for low field systems (Wu,
Chanmugam and Shaviv 1994) including IPs (Cropper et al. 1998).  It is
difficult to estimate the magnetic field of the white dwarf in V1432~Aql
owing to the poor quality of optical polarization data and lack of
cyclotron spectra.  Considering the analogy with another polar VV~Pup and
assuming that the peak in circular polarization of V1432~Aql corresponds to
the sixth to eighth harmonic of the cyclotron line, the magnetic field can
have a value in the range of 28--35 MG.  However, fixing $\epsilon_0$ to a
higher value also provides an equally good fit to the MOS spectra, and the
spectral fit is hardly sensitive to change in $\epsilon_0$.  The radius of
the accretion column is fixed to $10^8$ cm, close to its best fit value.
The specific accretion rate ($\dot{m}$) is also fixed to 1 g cm$^{-2}$
s$^{-1}$, which is characteristic of an intermediate accretion state.

Neutral absorbers with a full or partial covering of the source are used to
account for the intrinsic absorption seen in the spectra using the {\it
phabs} model of XSPEC and the H and He absorption cross-sections of
Balucinska-Church \& McCammon (1992) and Yan et al. (1998), respectively.
The multi-temperature plasma model and a partial covering absorption
component along with an absorption due to interstellar matter (ISM) and a
Gaussian line ({\it phabs$\times$pcfabs(sacg+gaussian)}) are used to fit
the time-averaged $RXTE$ spectra of V1432~Aql.  The two $RXTE$ spectra are
fitted simultaneously, since the spectral parameters from individual fits
are not significantly different.  The different intensities of the two
spectra are accounted for by different normalization constants.  The
multi-temperature plasma model does not predict the fluorescent emission
from cold Fe at 6.4 keV, although it does predict the emission lines at 6.7
and 7.0 keV from highly ionized Fe. A Gaussian line model centered at 6.4
keV with an assumed width of 0.1 keV is, therefore, added to fit the $RXTE$
spectra, and a reduced $\chi^2_{min}$ of 1.23 for 106 degrees of freedom
(dof) is obtained.  The best fit value for the column density of the
partially covering absorber is $N_H \simeq 28^{+5}_{-8}\times 10^{22}$
cm$^{-2}$ with a covering fraction of $55^{+5}_{-7}$\%.  The best fit value
for the mass of the white dwarf is found to be $1.21\pm0.10$ M$_{\odot}$.
Simultaneous fits to both the spectra along with the best fit model are
shown plotted in Figure~\ref{rxtespec} (top).  The bottom panel shows the
ratio of the data to the best fit model.  Based on these results the mass
of the white dwarf is fixed at 1.21 M$_{\odot}$ for the spectral analyses
of the $XMM$-$Newton$ data.

Since the $XMM$-$Newton$ data extend down to 0.2 keV, additional components
such as a blackbody, absorption due to the interstellar matter (ISM), and
an additional absorber fully covering the source are needed to fit and to
characterize the spectra.  The value of interstellar absorption along the
direction of V1432~Aql (distance=230 pc; Watson et al. 1995) is fixed as
follows: the ISM hydrogen column density search tool from the EUVE Web site
is used to estimate the column density in a given direction.  (Refer to
Fruscione et al. 1994 for the details of the database and the different
methods used by this tool to estimate the column density in a given
direction.)  For the 10 nearest neighboring sources at distances ranging
from 130 to 280 pc, the estimated column density varies over a large range
from $10^{19}$ to $1.6 \times 10^{21}$ cm$^{-2}$.  Schmidt \& Stockman
(2001) measured a visual extinction A$_v$=0.7$\pm$0.1 mag from the
interstellar absorption features using Hubble Space Telescope ($HST$) UV
spectrum for V1432~Aql.  The corresponding neutral hydrogen column density
estimated by them is $N_H\sim1.3 \times 10^{21}$ cm$^{-2}$.  $XMM$-$Newton$
spectra gave a poor value for the reduced $\chi^2$ (1.6 for 736 dof) with
the $N_H$ fixed at this value.  An independent measurement for the hydrogen
column density provided by the $ROSAT$ data (Staubert et al. 1994), gives a
value of $6.7^{+3.3}_{-2.2} \times 10^{20}$ cm$^{-2}$.  With the value of
$N_H$ fixed at the best fit value of the $ROSAT$ measurement and the two
extreme point values (68\% joint confidence limits for three parameters),
the $XMM$-$Newton$ data prefer to have the lowest value for the $N_H$ owing
to the ISM.  The lowest allowed value of $N_H$ from $ROSAT$ data can better
describe the $XMM$-$Newton$ data with $\Delta\chi^2\sim$180 as compared to
the $HST$ measurement.  Since the range of $N_H$ due to the interstellar
absorption obtained from the EUVE database is very large, (with the $ROSAT$
measured value well within the range), and the data indicate the lowest
value, a value of $4.5 \times 10^{20}$ cm$^{-2}$ is adopted for all
spectral analysis.

The two MOS data sets are initially fitted separately with a model
consisting of a blackbody component, an ISM component, a partial covering
absorber, and a multi-temperature plasma component, along with a Gaussian
line centered at 6.4 keV to account for the Fe fluorescent emission ({\it
phabs(bbody+pcfabs(sacg+gaussian))}).  In order to fix the value of the
line width in the Gaussian, the width is stepped over a range of 0.01--0.2
keV.  Values of the line width $>0.1$ keV begin to merge the line with the
continuum without significantly affecting the fit and the spectral
parameters.  Whereas, line widths $<0.05$ keV are narrower than observed.
Therefore, we fix the fluorescent line width at 0.05 keV.  The spectral
parameters derived from individual fits to two MOS spectra are found to
have similar values.  Therefore, in subsequent analyses simultaneous fits
to both MOS spectra are done to improve the statistics.  This model gives a
reduced $\chi^2_{min}$ of 1.5 for 737 dof and the best fit values for the
interesting spectral parameters are: a blackbody temperature kT$_{bb}
\simeq 73 \pm 2$ eV, a partial covering absorption density, $N_H$=$(8.0 \pm
1.0) \times 10^{22}$ cm$^{-2}$ with a covering fraction of $\sim68\pm1$\%,
and a Fe abundance of $0.83^{+0.17}_{-0.20}$ relative to solar.  The quoted
error bars (here and in rest of the paper) are for a 90\% confidence limit
for a single parameter (i.e. $\Delta \chi^2$=2.71).  The addition of an
extra absorber completely covering the source significantly improves the
fit by decreasing the reduced $\chi^2_{min}$ to 1.35 for 736 dof ({\it
phabs(bbody+phabs$\times$pcfabs(sacg+gaussian))}).  This model gives best
fit values of 88$\pm$2 eV for the blackbody temperature,
$N_{H1}$=1.7$\pm$0.3 $\times 10^{21}$ cm$^{-2}$ for the additional
absorber, and $N_{H2}$=$13.0^{+2.0}_{-1.0} \times 10^{22}$ cm$^{-2}$ for a
partial absorber covering $\sim65\pm1$\% of the source.  The Fe abundance
is 0.70$\pm$0.15 relative to solar (Anders \& Grevesse 1989).  The best fit
values of the spectral parameters obtained using the multi-temperature
plasma model are listed in Table 2.  The jointly fitted MOS spectra with
the best fit model are shown in Figure~\ref{xmmsacg} along with the
contributions from the individual model components.  The ratio of the data
to the best fit model are also shown to indicate the quality of the fit.
The inset of the figure shows an enlarged view of the 6.0 to 7.5 keV energy
range which contains the fluorescent line and the He-like and H-like Fe
K$_{\alpha}$ lines at 6.4, 6.7, and 6.95 keV, respectively.

The need for a strong blackbody component is checked by fitting the MOS
spectra using only absorbed multi-temperature plasma model and the Gaussian
line, but without the blackbody component.  The fit is unacceptable with a
reduced $\chi^2_{min}$ of $\sim$8 for 738 dof.  The residuals in terms of
the ratio of the data to the model are shown in Figure~\ref{softexcess}
(top).  The ratio plot shows a strong soft X-ray excess below 0.8 keV which
is well represented by a blackbody component as shown in
Fig.~\ref{xmmsacg}.

The line parameters of three Fe K$_{\alpha}$ lines are determined by
fitting a simple bremsstrahlung model along with three Gaussians in the
energy range of 5--9 keV.  The continuum around the line region is first
evaluated by fitting a bremsstrahlung model and ignoring the 6.0--7.2 keV
energy range containing the three lines.  Then, the ignored energy range
for the lines is retrieved and fitted with the three Gaussians.  The width
of each Gaussian is kept fixed at 0.05 keV.  The reduced $\chi^2_{min}$ is
1.05 for 243 dof.  This method gives the best fit values for the
fluorescent, He-like and H-like line energies as, $6.43^{+0.03}_{-0.02}$,
$6.70^{+0.03}_{-0.04}$, and $7.00^{+0.04}_{-0.03}$ keV, respectively; and
the line intensities as, 3.84 $\times 10^{-5}$, 3.07 $\times 10^{-5}$, and
2.55 $\times 10^{-5}$ photons cm$^{-2}$ s$^{-1}$, respectively.

\subsection{Photo-ionized Plasma Model}

The spectral model of Kinkhabwala et al. (2003) describes the X-ray
spectroscopic features expected from a steady-state, low-density
photo-ionized plasma and is used as an alternative model to the
multi-temperature plasma model.  The model has a simple geometry consisting
of a cone of ions irradiated by a point source located at the tip of the
cone.  The line emission is assumed to be unabsorbed and all lines are
assumed to be unsaturated at all column densities.  This model is available
as a local model `PHOTOION' for XSPEC.  The assumption of unsaturated lines
corresponds to the value of 7 for parameter 1 in the model and is
particularly suitable for use with spectra of CVs (see Mukai et al. 2003b).
This model contains a large number of free parameters including a power law
continuum, some system dependent parameters, and the ionic column densities
for various elements.  Some of the system dependent parameters are held
constant.  The covering fraction, defined as $f$=$\Omega$/4$\pi$ for a cone
geometry, is set at 0.5, since half of the X-ray flux impinges on the white
dwarf surface and the other half is radiated toward the observer.  The
quantity $\Omega$ is the solid angle subtended by the plasma.  The radial
velocity width $\sigma$ is held fixed to its default value of 100 km
s$^{-1}$.  The total power law luminosity is defined as
L(E)=AE$^{-\Gamma}$, where A is the normalization of the continuum and
$\Gamma$ is the power law index.  The distance to the source is fixed at
230 pc (Watson et al. 1995).

The MOS spectra are first fitted using the photo-ionized plasma model along
with ISM absorption in the energy range 0.8--10 keV, excluding the
blackbody dominated soft X-ray energy range ({\it
(phabs$\times$photoion)}).  The ISM column density is kept fixed at 4.5
$\times 10^{20}$ cm$^{-2}$ (see \S 4.1).  The individual ionic column
densities are not well determined owing to the moderate energy resolution.
We therefore set them to their default values of 0.0, except for the column
densities of the Fe ions.  The column densities of the 26 Fe ionization
states are varied one-by-one to find values that reproduce the line
emission in the 6--7 keV range, as well as the low energy ($<2$ keV)
continuum.  This exercise shows that several ions can be grouped together
to explain the Fe K-shell and L-shell emission.  The H-like and He-like Fe
K$_{\alpha}$ lines at 7.0 and 6.7 keV are respectively fitted with column
densities of $3 \times 10^{18}$ and $2 \times 10^{18}$ cm$^{-2}$.  As
mentioned in \S 3.1, the 6.4 keV fluorescent line shows a measurable
intrinsic width and can not be estimated by varying the neutral Fe column
density alone.  It is, therefore, necessary to adjust the column densities
of the remaining Fe ions (Fe I--XXIV) in three different groups: Fe I--XVI,
Fe XVII--XIX, and Fe XX--XXIV with the estimated column densities of $8
\times 10^{18}$ cm$^{-2}$, $3 \times 10^{18}$ cm$^{-2}$, and $3 \times
10^{17}$ cm$^{-2}$, respectively, for a good fit to the data.  With the
ionic column densities for Fe fixed, the best fit value of the power law
index is $\Gamma$=0.59, and the reduced $\chi^2_{min}$ with 345 dof is
1.16.  The high values of the estimated ionic column densities for Fe
I--XIX ions should at most be taken as upper limits until higher resolution
spectra become available.

After achieving a good fit to MOS spectra in the energy range of 0.8--10
keV, the soft X-ray (0.2--0.8 keV) energy range is included in the
analysis.  The residuals show a large excess below 0.8 keV and the reduced
$\chi^2_{min}$ is unacceptably high at $\sim$10 for 742 dof (see
Figure~\ref{softexcess}, bottom).  The addition of a blackbody component
significantly improves the fit, decreasing the reduced $\chi^2_{min}$ to
1.28 for 740 dof ({\it phabs(bbody+photoion)}).  The ionic column densities
are held fixed at the above values.  However, the power law index $\Gamma$
is allowed to vary and is found to be 0.64.  The best fit value for the
blackbody temperature is $80^{+1}_{-2}$ eV, similar to that for the
multi-temperature plasma model.  The best fit parameters for the
photo-ionized plasma model are listed in Table 3.  A joint fit of the model
to the MOS spectra, showing the contribution from all the components, is
shown plotted in Figure~\ref{xmmphoto} (top).  The bottom panel shows the
residuals and the inset shows an enlarged view of the Fe K$_{\alpha}$ line
complex.  Thus, both spectral models that have been used here, describe the
moderate resolution X-ray spectra equally well for V1432~Aql, and hence, it
is difficult to distinguish between them using the available X-ray data.

\section{Spin Phase Resolved X-ray Spectroscopy}

To gain a better insight into the spectral behavior of the source as a
function of the spin phase, we analyze the spectra extracted for six
unequal phase intervals, i.e., $\phi$=0.18--0.30, $\phi$=0.30-0.40,
$\phi$=0.40--0.54, $\phi$=0.54--0.73, $\phi$=0.73--0.90 and
$\phi$=0.90--1.18.  This division ensures that spectra in each phase
interval have sufficient counts to provide reliable values of different
spectral parameters.  Since the use of model B requires a higher resolution
data to provide an accurate information about the line parameters, and
model A (see \S 4.1) along with other model components as employed to
describe the average spectra provides a good description of an accretion
column in a polar, we have used model A to fit the phase-resolved spectra.
During the fit to phase resolved spectra the black body temperature is kept
fixed at 88 eV obtained from the phase averaged spectral fit.  Allowing the
black body temperature to vary adjusts the value of other parameters like
absorbing column densities and the covering fraction and gives only a
slightly better fit with $\Delta \chi^2$=9, 13, 16, 17, 55, 22, and 20,
respectively for each of the phase intervals listed in Table 2.  However,
variation of the black body temperature with the spin phase is not very
likely.  The jointly fitted MOS spectra corresponding to all six phase
intervals, along with the individual best fit model components are plotted
in Figure~\ref{pha_res_spec}.  The values of the best fit spectral
parameters obtained from the phase resolved spectral analysis are also
listed in Table 2.  The degrees of freedom are different for each spectrum,
because the binning procedure requires a minimum of 20 counts per bin.
Several spectral parameters show variations over the spin cycle (see Table
2), and are shown plotted in Figure~\ref{specpara} as a function of the
spin phase.  Variations in the average intensity, the absorption
components: $N_{H1}$, the column density due to an absorber fully covering
the source; $N_{H2}$, the column density due to an absorber partially
covering the source; the covering fraction $C_f$ associated with $N_{H2}$,
and the 6.4 keV line flux are shown.  The $N_{H1}$ value remains almost
constant at $\sim$$2.4 \times 10^{21}$ cm$^{-2}$ during phase 0.18--0.73
except for $\phi$=0.3--0.4 during which it is at a lower value of $0.9
\times 10^{21}$ cm$^{-2}$.  The $N_{H2}$ value remains low and roughly
constant at an average value of $\sim$ $1.1 \times 10^{23}$ cm$^{-2}$
during most of the spin cycle (i.e.  $\phi$=0.18--0.90) but shows a very
high value of $7.6^{+3.0}_{-4.1}$ $\times 10^{23}$ cm$^{-2}$ during the
last spin phase interval of 0.90--1.18.  The $C_f$ value is the same within
the error bars during the phase range 0.9--1.54 but is anti-correlated with
the intensity during the rest of the spin cycle.  The variation in the line
flux is correlated with the average change in the source intensity.

\section{Discussion}

\subsection{Power Spectra and the Spin Period}
Several prominent peaks at frequencies corresponding to various
periods related to the system have been identified.
A peak at a period of 12150.72$\pm$1.14~s has been observed in the $ROSAT$
power spectra (Fig.~\ref{rosatpds}). This value
agrees well with the determinations of the
spin period ($\omega$) based on X-ray and optical photometry of V1432~Aql as
reported by Friedrich et al. (1996) (12150.7$\pm$0.5~s based on the
$\chi^2$ period search method), Staubert et al. (2003)
(12150.328$\pm$0.045~s), and Mukai et al. (2003a) (12150.432~s).  With
this presumed spin period, the
dominant peak in the $XMM$-$Newton$ power density spectra can be
interpreted as due to the second harmonic of the spin period.  The
circular polarization data show a dominant peak at the
one day alias of the first harmonic of $\omega$ in the power
density spectrum.  Smaller peaks at the fundamental frequency and
higher harmonics are also visible in the polarization data.  Thus the
optical polarization is consistent with the presence of the
fundamental period of 12150~s.

\begin{figure}
\epsscale{0.8}
\plotone{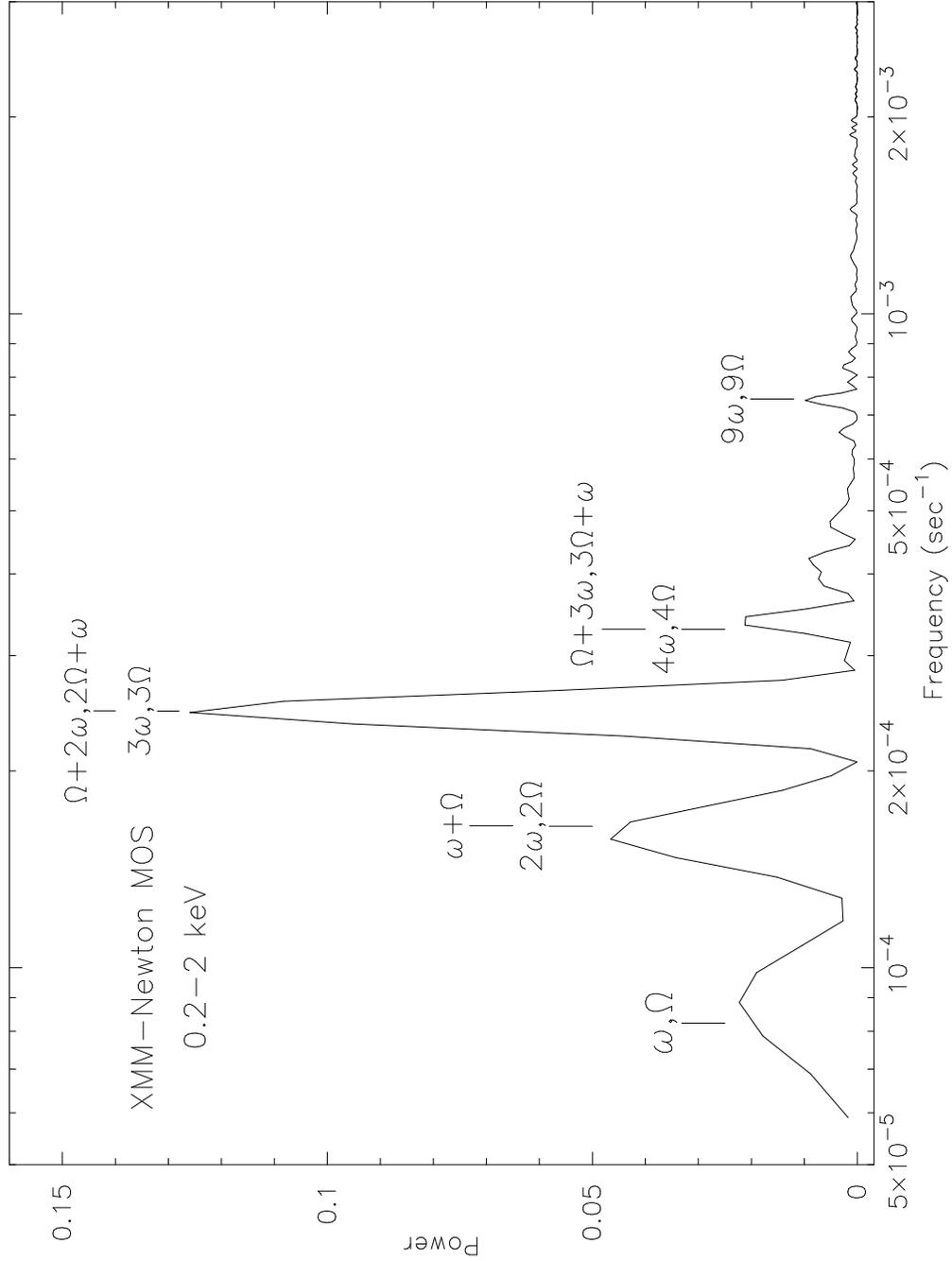}
\caption{Power density spectrum of V1432~Aql obtained from a long
continuous $XMM$-$Newton$ observation in the 0.2--2 keV energy band.
Various identified components are marked.
\label{xmmpds}}
\end{figure}

\begin{figure}
\plotone{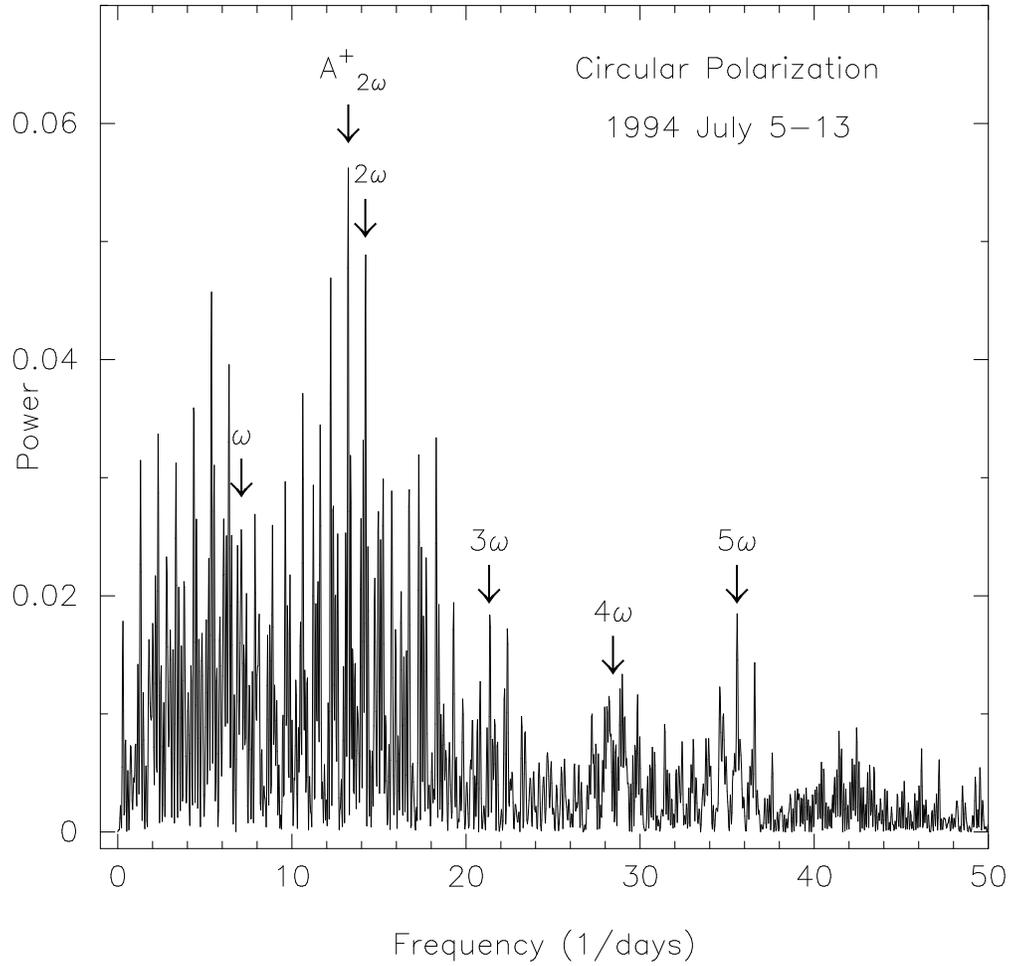}
\caption{Power density spectrum obtained using discrete Fourier
transform of optical circular polarization data taken from $SAAO$
during 1994 July 5--13.  The expected positions of frequency
components corresponding to spin period, and its various harmonics are
marked along with a one day positive alias of 2$\omega$ component.
\label{circpds}}
\end{figure}

\begin{figure}
\epsscale{0.8}
\plotone{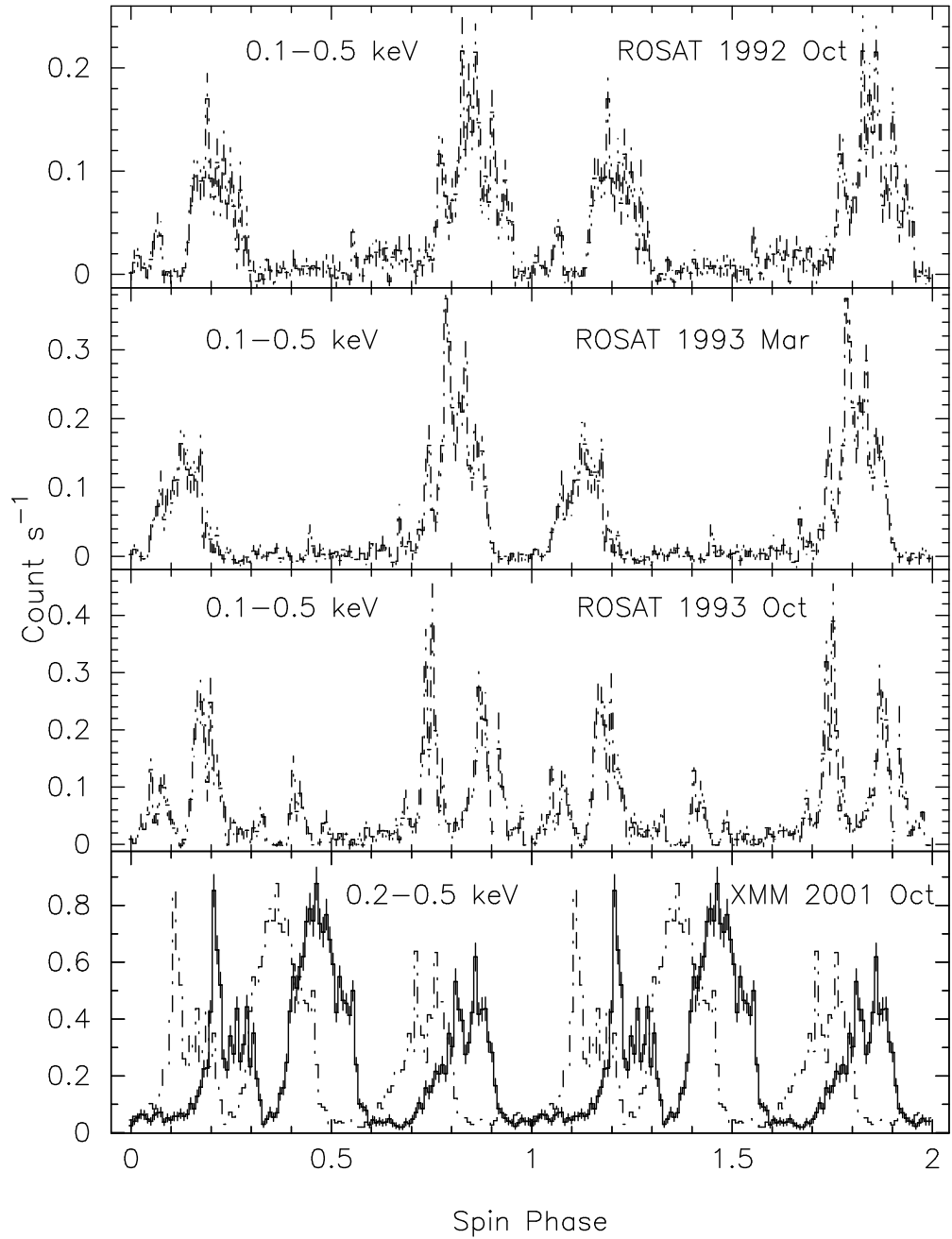}
\caption{Spin folded X-ray light curve from the $ROSAT$ (first three
panels; 0.1--0.5 keV energy band) and the $XMM$-$Newton$ (bottom
panel; 0.2--0.5 keV energy band) observations (see Table 1).  For
folding the data, we have used the ephemerides for the spin period as
given by Staubert et al. (2003) (dash-dotted line) and Mukai et
al. (2003a) (solid line curve in bottom panel).  The bin time is
$\sim$100~s throughout. Two phase cycles are shown for clarity.
\label{softlc}}
\end{figure}

\begin{figure}
\epsscale{0.8}
\plotone{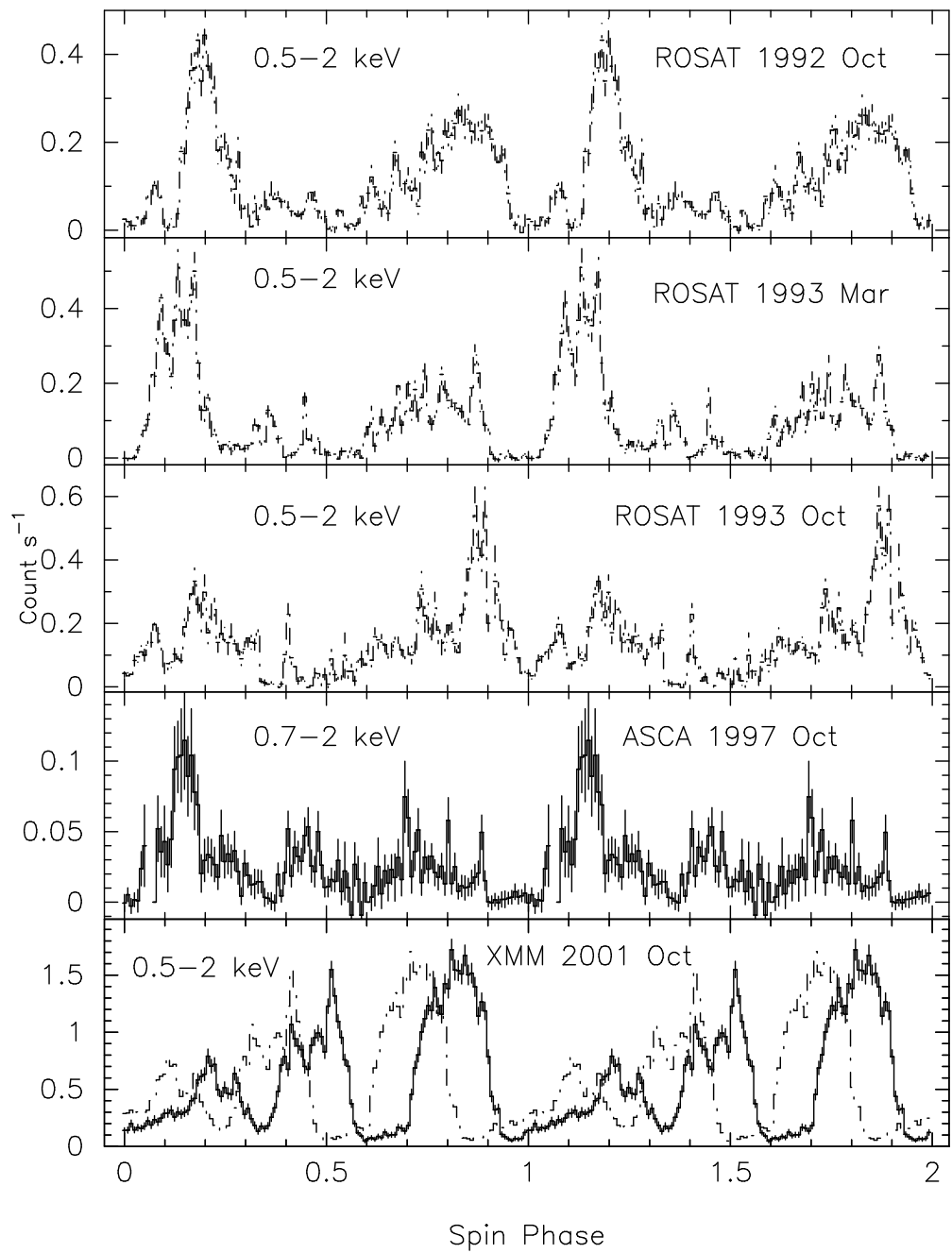}
\caption{Spin folded X-ray light curves from the $ROSAT$ (first three
panels; 0.5--2 keV energy band), the $ASCA$ (panel four; 0.7--2 keV
energy band) and the $XMM$-$Newton$ (bottom panel; 0.5--2 keV energy
band) observations.  Same ephemerides with the same line styles, as
used in Fig.~\ref{softlc}, are used here for folding the data.
The bin time is $\sim$100~s throughout.
\label{medlc}}
\end{figure}

\begin{figure}
\epsscale{0.8}
\plotone{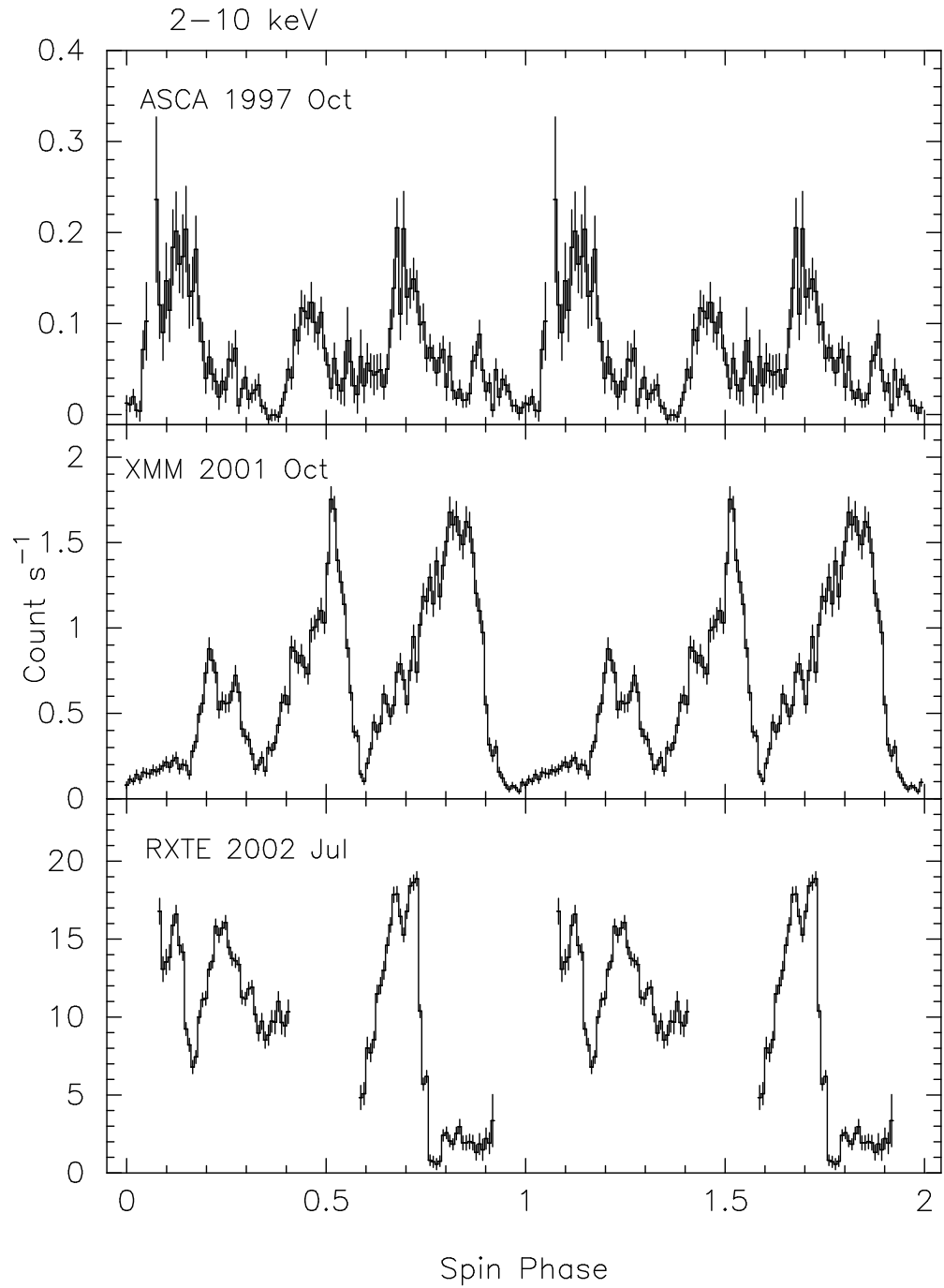}
\caption{Hard X-ray spin folded light curves in the 2--10 keV energy band
from the $ASCA$, the $XMM$-$Newton$ and the $RXTE$ observations.  For
folding the data, the spin ephemeris as given by Mukai et al. (2003a)
has been used.  The bin time is $\sim$100~s.
\label{hardlc}}
\end{figure}

\begin{figure}
\epsscale{0.8}
\plotone{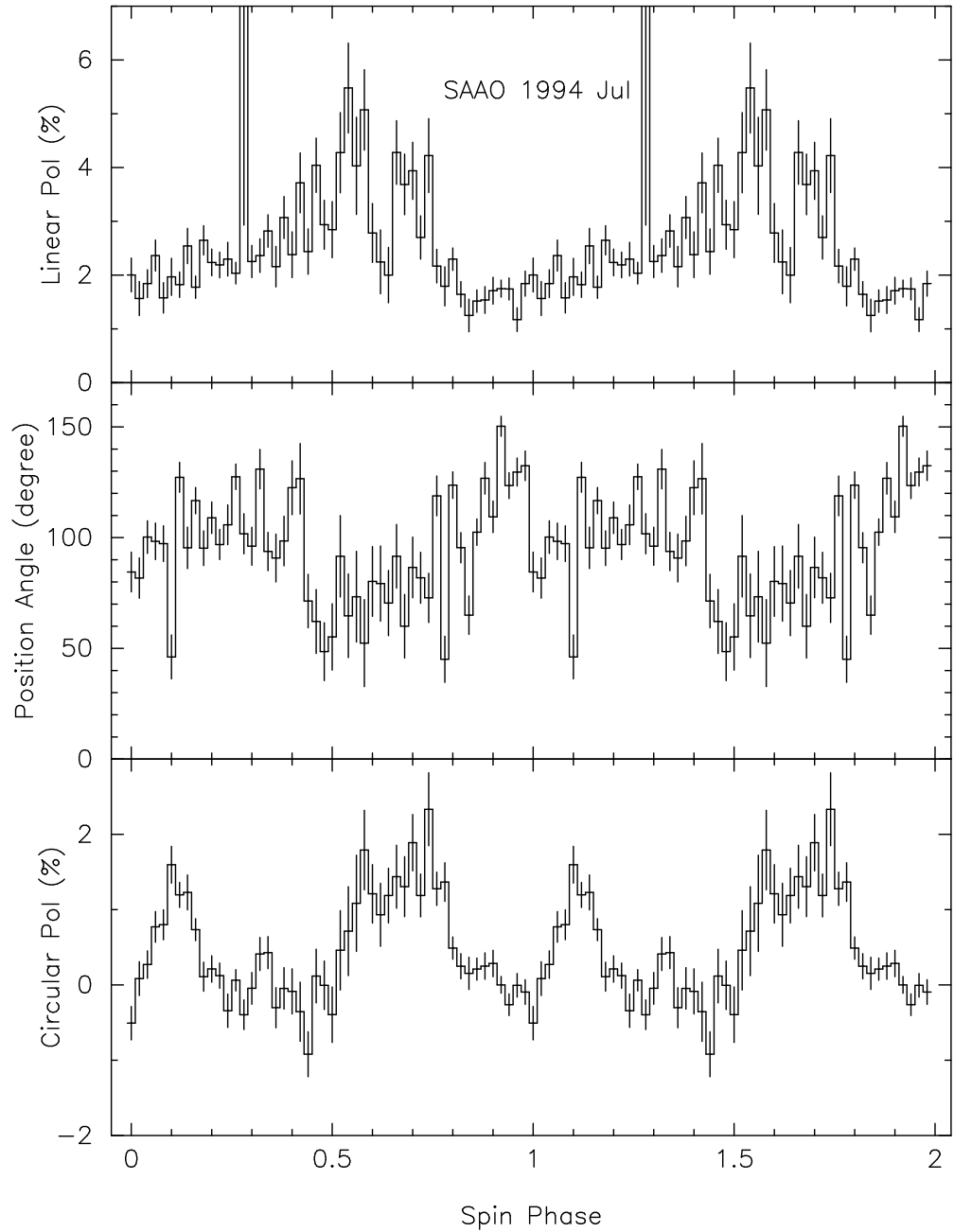}
\caption{Spin folded polarization data in the optical band obtained from
the $SAAO$ during 1994 July 5--13. The data have been folded using the
spin ephemeris as given by Staubert et al. (2003).  The bin time is
$\sim$243~s.
\label{optlc}}
\end{figure}

\begin{figure}
\epsscale{0.9}
\plotone{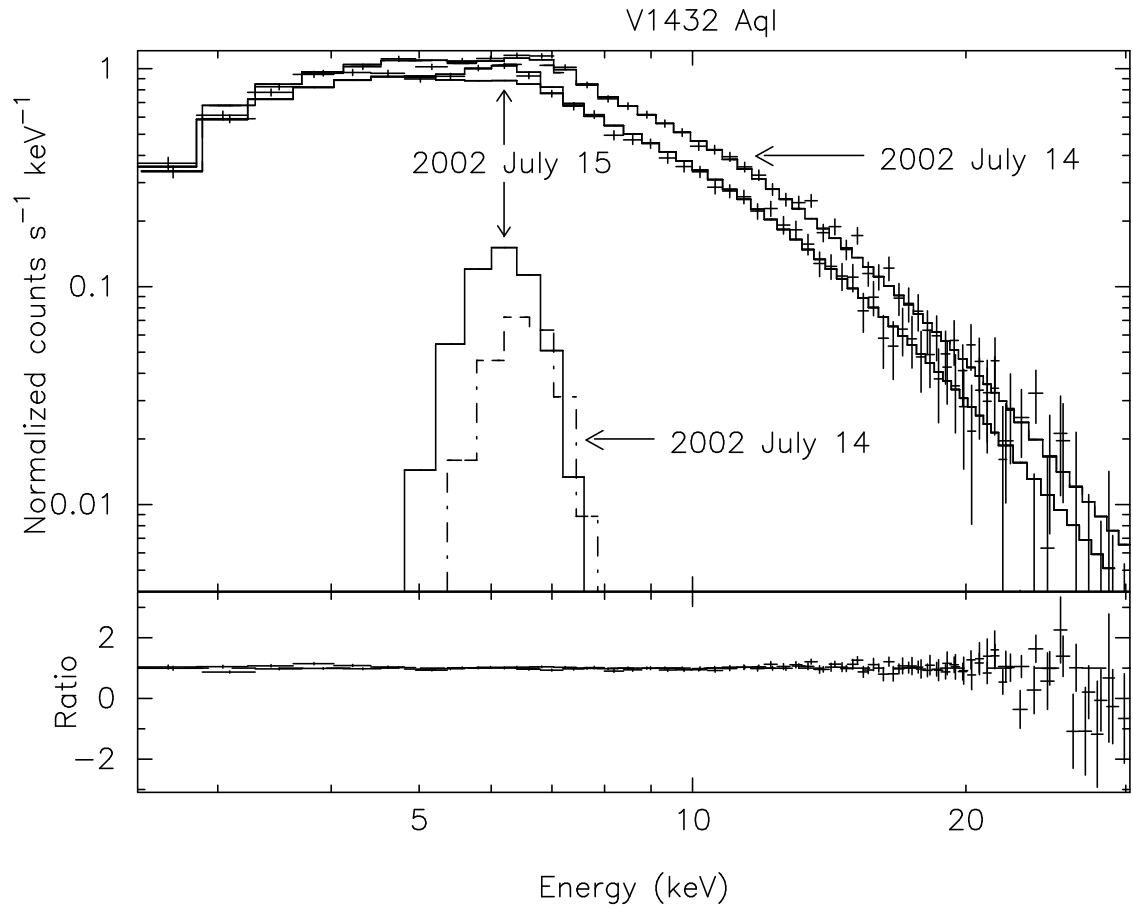}
\caption{$RXTE$ PCA spectra and the best fit multi-temperature
plasma model with a Gaussian line component (top), and the ratio
of the data to the best fit model (bottom).
\label{rxtespec}}
\end{figure}

\begin{figure}
\plotone{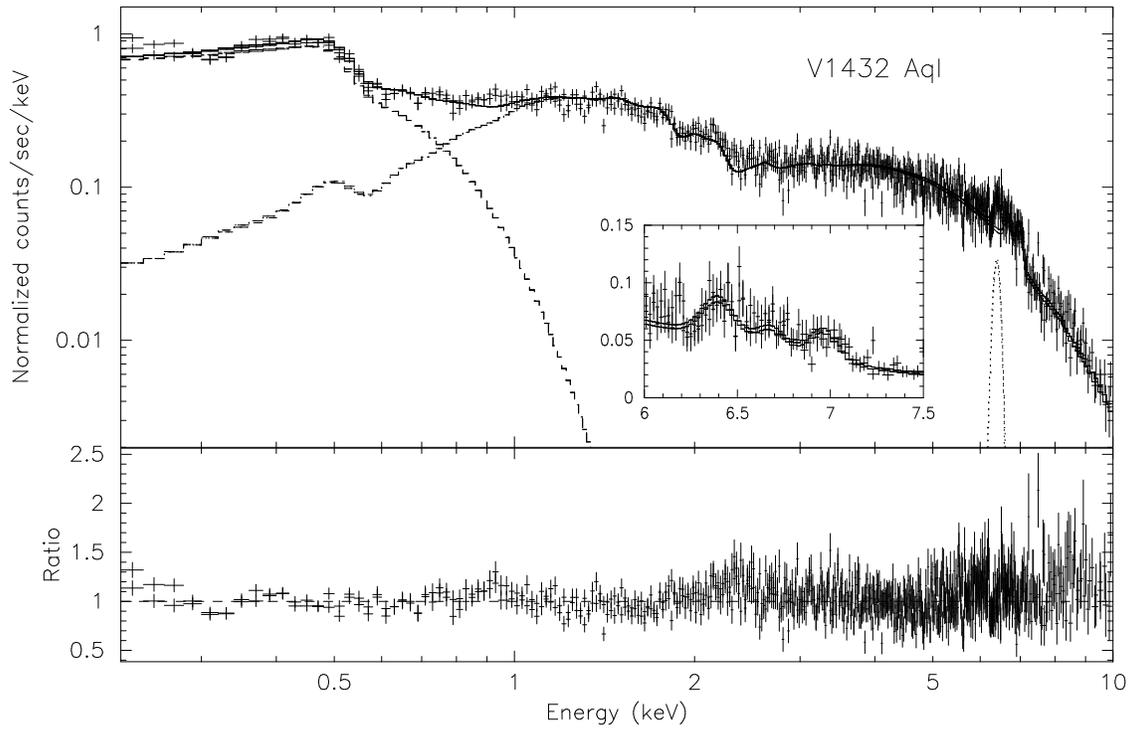}
\caption{The EPIC MOS1 and MOS2 spectra and the best fit model.  The
top panel shows data fitted with a best fit model consisting of a
blackbody, an absorbed multi-temperature plasma (Cropper et al. 1999),
and a Gaussian emission line.  The additive components are shown
individually.  The bottom panel shows the ratio of the data to the
best fit model and the inset shows a magnified view of the energy
range containing the Fe K$_{\alpha}$ line complex.
\label{xmmsacg}}
\end{figure}

\begin{figure}
\plotone{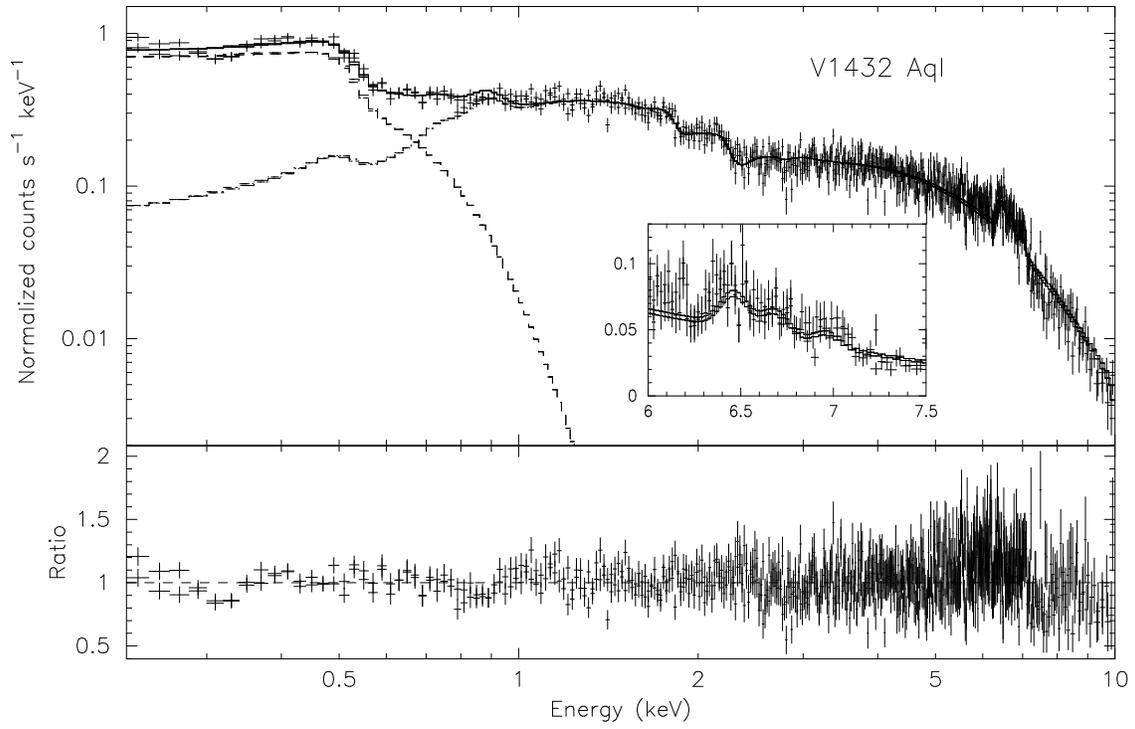}
\caption{Same as in Fig.~\ref{xmmsacg}, but for the photo-ionized
plasma model. \label{xmmphoto}}
\end{figure}

\begin{figure}
\plotone{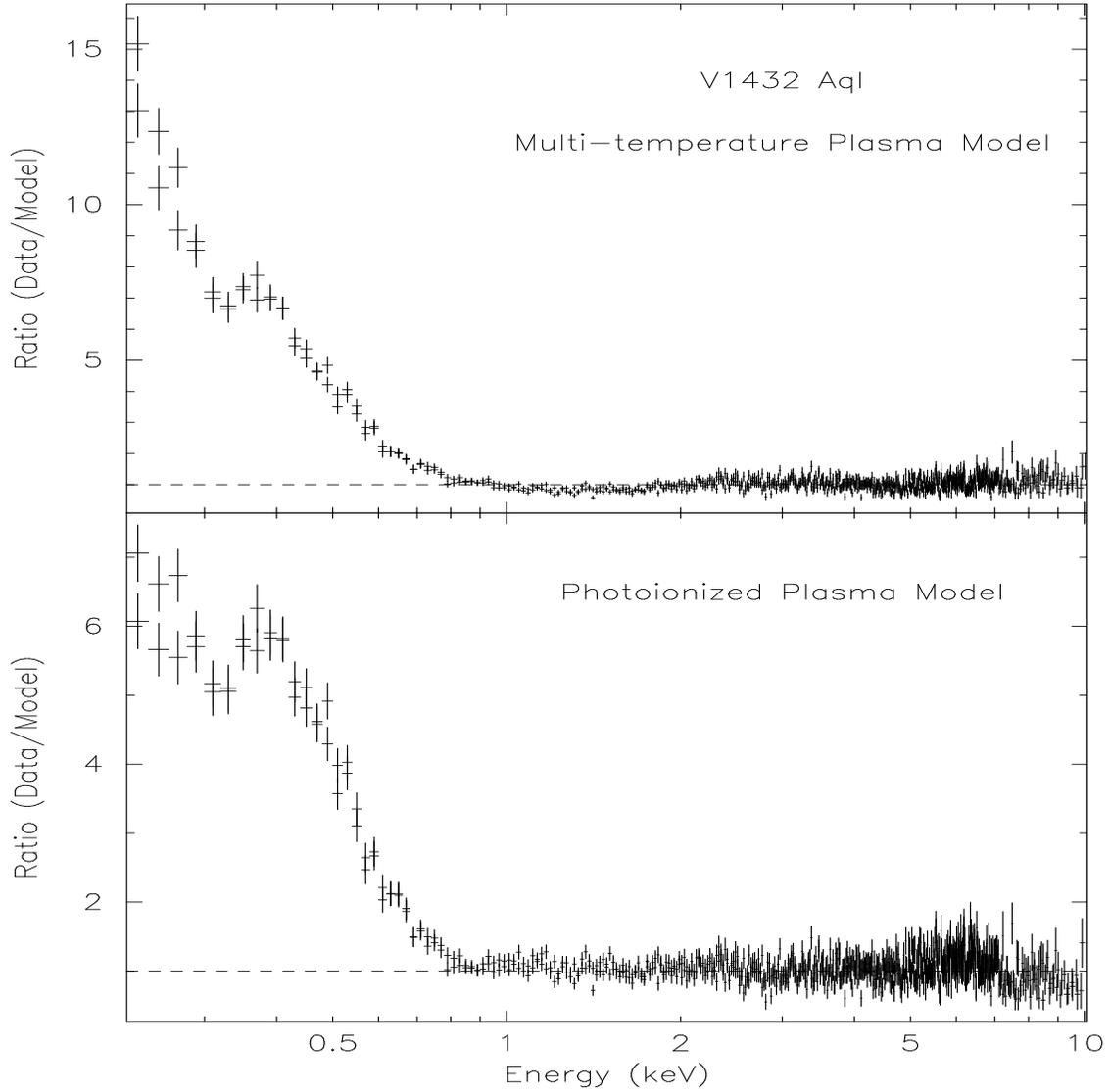}
\caption{Ratio of the data to the multi-temperature
plasma+Gaussian line model without a blackbody component for the
average MOS spectra in the energy range 0.2--10 keV (top) and for
the photo-ionized plasma model without blackbody component (bottom).
A clear excess in soft X-rays (below 0.8 keV) is visible in
both cases.
\label{softexcess}}
\end{figure}

\begin{figure}
\epsscale{.80}
\plotone{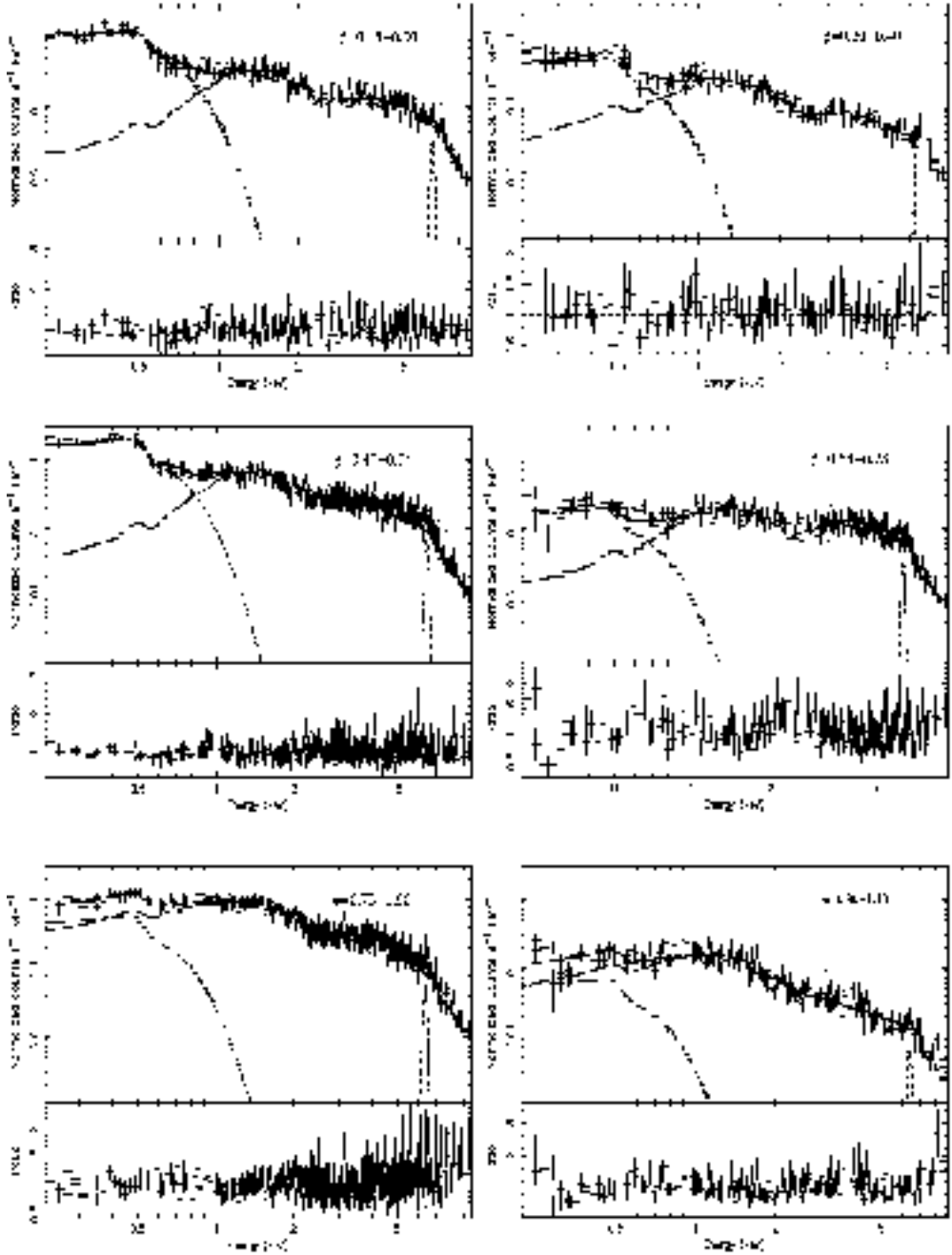}
\caption{Phase resolved MOS spectra of V1432~Aql corresponding to the
six unequal intervals of the spin cycle, along with the individual
components of the best fit model (top panels). The ratio of the data
to the best fit model for each spectra is plotted in the bottom
panels.  The corresponding phase ranges are labeled for each spectra.
\label{pha_res_spec}}
\end{figure}

\begin{figure}
\epsscale{0.8}
\plotone{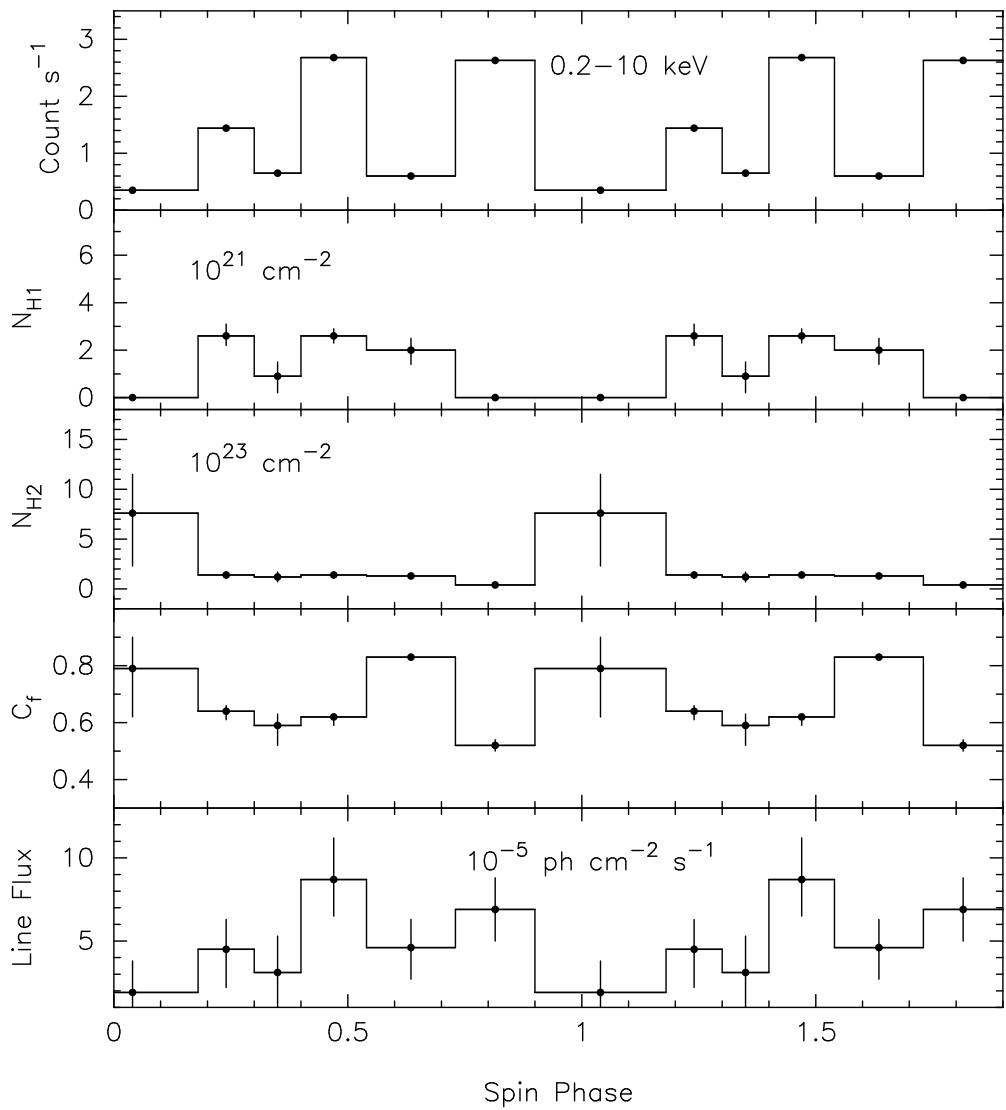}
\caption{Variation of spectral parameters (see text) as a function of
spin phase. The intensity variation is shown in the top panel.  The
second panel from top shows change in the column density $N_{H1}$ due to
an absorber fully covering the source.  Variations in the column
density, $N_{H2}$, due to a partial covering of the source and the
associated covering fraction are plotted in the third and fourth panels
from the top, respectively.  The bottom panel represents a change in
the flux of the Fe 6.4 keV line. Errors plotted are with 90\%
confidence.
\label{specpara}}
\end{figure}

The power spectra for the $ROSAT$ medium energy bands also
show a peak at the spin frequency $\omega$, but with a relatively
smaller amplitude as compared to the total energy band power spectrum.
The X-ray power spectra are
strongly dependent on some of the geometrical parameters of the
system, e.g. the inclination angle ($i$) and the magnetic co-latitude
($\beta$).  In addition, for asynchronous polars, there will also be
an effect from its non-synchronous rotation, since the accretion
stream interacts with different field lines as the white dwarf
rotates.  Moreover, if the time interval over which the power spectrum
is calculated, is reasonably longer than the beat period of the
system, then the power spectrum is expected to be dominated by side
band frequencies of the fundamental periods instead of the fundamental
periods themselves.  According to Wynn \& King (1992) the power
spectra of asynchronous polars with high values of $i$ and $\beta$,
show strong peaks at 2$\omega$-$\Omega$ and $\Omega$ components.  The
2$\omega$-$\Omega$ component is prominent in the power spectrum from
the medium energy $ROSAT$ data, whereas it has relatively smaller power in
the power spectrum of the total energy band.  This component is merged
with the $\omega$ and $\Omega$ frequency components in the power
spectrum from $XMM$-$Newton$ data (Fig.~\ref{xmmpds}).
This side band component
was also observed in the power spectra of the two other asynchronous
polars BY~Cam (Mason et al. 1998) and RX J2115-5840 (Ramsay et
al. 1999) in the optical data.  The presence of this component
suggests a diskless accretion geometry for asynchronously rotating
MCVs (Wynn \& King 1992).

Although there are three clear humps during the 12150~s period in the
$XMM$-$Newton$ and $ASCA$ light curves, the repetitive appearance of these
features is preserved with roughly the same time structure (see Fig.1 of
Paper 1).  This recurrence of the features in the light curve dominates
over the 12150~s periodicity as the true spin period, leading to a dominant
peak in the power spectrum at the frequency corresponding to the second
harmonic of the spin period (i.e.  3$\omega$ component at $\sim$4050~s; see
Fig.~\ref{xmmpds}) rather than at the true spin period of 12150~s.  This
kind of intensity variation and the appearance of the dominant power at
$\sim$4050~s period led Mukai (1998) (and Singh \& Rana 2003) to suggest
that V1432~Aql may be an IP with a $\sim$4050~s spin period.  However, the
present analyses of X-ray timing data, spectra, and optical polarization
data are consistent with V1432~Aql being an asynchronous polar and not an
IP.

\subsection{Possible Accretion Geometries For V1432~Aql}

V1432~Aql shows varying light curve profiles with a double hump structure
during the $ROSAT$ observations in 1992--1993 changing to a triple hump
structure during the $ASCA$ and $XMM$-$Newton$ observations in 1997 and
2001.  The circular polarization data taken in 1994 shows a double humped
profile bearing a striking resemblance to the soft X-ray $ROSAT$ light
curves.  The occurrence of both the negative and the positive value of the
circular polarization strongly suggests the presence of two sites of
polarization that are viewed alternately, indicating the presence of two
magnetic poles in V1432~Aql.

Staubert et al. (2003) showed that the accretion process in V1432~Aql can
be described using a dipole accretion model (also see Geckeler \& Staubert
1997), and fitted their model to the optical and the X-ray data from
$ROSAT$ that show a double hump profile.  The appearance of the two broad
humps in the $ROSAT$ soft X-ray light curves results when the X-ray
emitting regions are in direct view crossing the line-of-sight alternately.
In between the appearance of the humps the emitting regions are out of view
or viewed through absorbing material that is almost opaque to X-rays below
0.5 keV.  On the other hand, the 0.5--2 keV X-ray light curves during the
rest of the spin cycle show a comparatively higher and variable intensity,
indicating that X-rays above 0.5 keV manage to escape from the absorbing
material.  The observed separation of the $\sim$0.65 spin phase between the
two humps (in the first three panels of Fig.~\ref{softlc}) suggests that
the two accretion foot prints are separated by $\sim$234$^{\circ}$.  Thus,
the accretion via two poles with the two hot spots that are not anti-podal
can explain the intensity variation in the $ROSAT$ light curves.  During
the $ROSAT$ 1993 October observations each of the two humps representing
the two accretion spots appear to split into two owing to the presence of
two dips (at $\phi$=0.12 and 0.8).

The triple hump profiles of the $ASCA$ and $XMM$-$Newton$ data are
difficult to explain using the two pole accretion model with two
off-centered hot spots.  To accommodate them in a two pole accretor model,
the two nearby humps around spin phases 0.2 and 0.45 in $XMM$-$Newton$
light curves, could represent one dominant pole that is split into two
because of the presence of a narrow dip and the other hump at $\phi$=0.8
represent the second pole.  This dip could be due to absorption by the
accretion stream or a split or double stream system.  If the two close
peaks represent one dominant pole, then this implies that the pole remains
in view for almost half the spin cycle, whereas the $ROSAT$ data show that
the same pole is only visible for 0.2 of the spin phase.  It is not clear
why the dominant pole remains in view for a longer time during the $ASCA$
and the $XMM$-$Newton$ observations.  The dip at the $\sim$0.35 spin phase
is equally prominent in all the energy bands showing little X-ray emission.
Therefore, the possibility of a split or double stream system is more
likely the cause for the appearance of the triple hump in the $ASCA$ and
$XMM$-$Newton$ data, and not the absorption.  The phase resolved
spectroscopy presented in \S 5 (Fig.~\ref{specpara}) also favors this
interpretation as no extra absorption is seen during this phase range.  The
second spot is visible during $\phi$=0.7--0.9 when the first goes behind
the surface of the white dwarf.  The location and the extent of the second
hot spot in the $ASCA$ and the $XMM$-$Newton$ are very similar to that in
the $ROSAT$ data.  The low intensity during $\phi$=0.6--0.7 and 0.90--1.18
is associated with a high absorption and a large covering fraction by a
very thick absorber.  The highest intensity peak centered at phase 0.8 has
the smallest covering fraction of the source by an absorber, and gives the
least hindered view toward the accreting pole.

The spin folded circular polarization data support a complex magnetic field
geometry with three hot spots.  The two positive humps observed in the
circular polarization data (near phase 0.1 and 0.7) indicate the same
magnetic polarity for the two emission regions and a negative polarization
near phase 0.45 suggests another hot spot with opposite polarity.  In this
scenario, the two humps at phase 0.25 and 0.85 in the $XMM$-$Newton$ light
curve represent two hot spots with the same polarity and the third at phase
0.50 represents a region with opposite polarity.  The different amplitudes
of the two positive humps in the polarization suggest different magnetic
field strengths for the two poles with the same polarity.  The peak at
phase 0.7 has relatively higher polarization, suggesting a higher magnetic
field strength and hence lower accretion on that pole (see Stockman 1988
and Wickramasinghe \& Ferrario 2000), resulting in lower X-ray emission
near phase 0.85 (a small phase shift is likely due to two different
ephemerides used).  The other pole with a relatively smaller magnetic field
corresponds to higher X-ray intensity (near phase 0.25). The hot spot with
opposite polarity shows a smaller circular polarization and hence
corresponds to the higher X-ray intensity (near phase 0.50).  The
separation between the three humps in $XMM$ data suggests that the first
two spots are about 90$^{\circ}$ apart and the second and the third spots
are further $\sim$126$^{\circ}$ apart in longitude.  Thus, the recent X-ray
and circular polarization data suggest a field geometry with at least three
hot spots on the surface of the white dwarf.

A similar variation in the circular polarization data is also observed in
the polar EF~Eri (Piirola, Reiz, \& Coyne 1987).  Meggitt \& Wickramasinghe
(1989) proposed a dominant quadrupole field structure with four active
spots on the surface of the white dwarf for EF~Eri using optical
polarimetric data.  Mason et al. (1995) provided evidence for the presence
of multi-pole magnetic field structure in another asynchronous polar
BY~Cam.  Thus, the presence of more than two accretion spots on the surface
of the white dwarf is not an uncommon characteristic of polars.

The difference in the morphology of X-ray light curves from the $ROSAT$ and
$XMM$-$Newton$ (or $ASCA$) data can well be due to the beat period.  If we
choose the zero of the beat phase at the starting time of the $ROSAT$ 1992
observation and a beat period of 49.8955 days, then the beat phases for the
start time of the $ROSAT$ 1993 March and October, optical polarization
data, $ASCA$ 1997 October, $XMM$-$Newton$ 2001 October, and $RXTE$ 2002
July 14 and 15 are 0.48, 0.42, 0.73, 0.98, 0.88, 0.46 and 0.48,
respectively.  If the different X-ray profiles are due to the different
beat phases of the observations then the transition from the double hump to
the triple hump profile most likely occurs between beat phases 0.5--0.7.
This conclusion is not affected even if we consider the spin up of the
white dwarf in estimating the beat phases.

\subsection{X-ray Spectrum}

The low and medium resolution X-ray spectra of V1432~Aql obtained with the
$RXTE$ and the $XMM$-$Newton$ can be reproduced using either a
multi-temperature plasma model with multiple absorbers or a
photo-ionization model.  The phase-averaged spectra from the $RXTE$ extend
to $\sim$30 keV and can be modeled by a simple bremsstrahlung with
kT$\geqslant$90 keV (90\% confidence limit).  This temperature is among the
highest seen in MCVs.  The use of a multi-temperature plasma model for the
$RXTE$ data gives an estimated mass of the white dwarf of 1.2$\pm$0.1
M$_{\odot}$.  This is similar to that of PQ~Gem, a soft IP, for which James
et al. (2002) report a mass of $1.21^{+0.07}_{-0.05}$ M$_{\odot}$ using the
$RXTE$ data and the same multi-temperature plasma model.  The high mass
estimate of the white dwarf is consistent with the idea that V1432~Aql
might have undergone a nova explosion in not too distant past.  This is
also consistent with the observed asynchronism in the system.  The medium
resolution $XMM$-$Newton$ spectra can be reproduced using this mass
estimate in the multi-temperature plasma model.  The multi-temperature
plasma produced in the post-shock region must, however, be viewed through
two absorbers -- a thin absorber of (1.7$\pm$0.3) $\times 10^{21}$
cm$^{-2}$ covering the entire source and a thick absorber of
($1.3^{+0.2}_{-0.1}$) $\times 10^{23}$ cm$^{-2}$ covering 65\% of the hard
X-ray emission.  Since the plasma emission originates in the post-shock
region, these absorbers are most probably situated in front of the
post-shock region or in the pre-shock flow.  The abundance of Fe in the
plasma is found to be nearly solar (except during the lowest intensity
phase) and provides a good fit to the emission lines at 6.7 and 7.0 keV
(see inset in Fig.~\ref{xmmsacg}).

Alternatively, a photo-ionization model can also reproduce the MOS spectra.
Photo-ionization models have recently been successful in explaining the
high resolution $Chandra$ HETG spectra of IPs, and in fact might be
preferred over the multi-temperature plasma models like those used for
cooling flows (Mukai et al. 2003b).  Although the photo-ionization model
has a slight edge (in terms of $\chi^2_{min}$) over the multi-temperature
plasma models used here, higher resolution grating spectra of V1432~Aql are
required to decide between the two models.  The important differences
between the two models are seen in excitation lines (e.g., N VI, O
VII-VIII, Ne IX-X, Mg XI-XII, Si XIII-XIV, and Fe L-shell emission lines).
Therefore higher resolution data are important to resolve the line emission
from these elements that occur at low energies (below $\sim$2 keV; Mukai et
al. 2003b).  However, the photo-ionization model can successfully reproduce
the three Fe line components at 6.4, 6.7, and 7.0 keV without requiring a
separate Gaussian component for modeling the line at 6.4 keV, as with the
multi-temperature plasma model, but it requires extremely large ionic
column densities for the Fe ions (see Table 3).  The 6.7 keV and 7.0 keV
lines provided a reasonably good fix for the column densities of Fe XXV and
Fe XXVI ions.  The strong line emission observed at 6.4 keV can only be
reproduced by having a large column for Fe I--XIX ions.  Lack of resolution
in the present spectra, does not allow us to derive the individual column
densities of all the ions in the photo-ionized regions.  The ionic column
densities derived for Fe I--XIX ions are tentative and likely to be
overestimates.  In addition, the use of the input power law continuum is
not well known and the ionic column densities for Fe are adjusted to fit
the observed lines.

The thermal plasma models (a simple bremsstrahlung or multi-temperature
plasma) require an additional Gaussian component at 6.4 keV, which is
presumably due to fluorescent emission from (cold or warm) Fe.  The fit
with the multi-temperature plasma model shows that the Fe line at 6.4 keV
has some intrinsic line width indicating the presence of warm (partially
ionized) Fe in addition to cold (or un-ionized) Fe.  Alternatively, the
width of 6.4 keV line can also be due to the velocity broadening but
present data can not distinguish between these two mechanisms.  The
photo-ionized plasma model supports this trend and explicitly demonstrates
the need for high column densities for Fe I--XIX.  The variations in Fe 6.4
keV line flux closely resemble the variations in the average X-ray
intensity in 0.2--10 keV energy band, further indicating that the spin
period is 12150~s.

\subsection{White Dwarf Mass Determination Using Fe K$_{\alpha}$ Lines}

The mass of the white dwarf can also be estimated using the intensity ratio
of K$_{\alpha}$ emission lines from abundant heavy elements, as described
by Fujimoto \& Ishida (1997).  For a plasma in collisional ionization
equilibrium (CIE), the intensity ratio of H to He-like K$_{\alpha}$ lines
is a function of the plasma temperature (Mewe et al. 1985).  The plasma
temperature is directly related to the gravitational potential of the white
dwarf.  Therefore, the mass of the white dwarf can be estimated using the
mass-radius relationship (Nauenberg 1972), if the plasma temperature is
known.  Here we use only the Fe K$_{\alpha}$ lines as lines from other
element are not resolved.  This method is independent of the uncertainties
in geometrical parameters like inclination angle and the mass-radius
relationship of the secondary star.  The intensity ratio of H to He-like Fe
K$_{\alpha}$ line is $\sim$0.83 for V1432~Aql (see \S 4.1).  A similar
value ($\sim$0.9) for the line intensity ratio of Fe K$_{\alpha}$ lines is
reported by Fujimoto (1996) for the IP, V1223~Sgr.  This corresponds to a
lower limit of $\sim$38~keV for the shock temperature and a lower limit of
0.82M$_{\odot}$ for the white dwarf mass in V1223~Sgr.  V1432~Aql shows a
very hard X-ray spectrum and similar line intensity ratio for the H to
He-like Fe K$_{\alpha}$ lines suggesting a similar lower limit for the mass
of the white dwarf that is consistent with the higher mass derived from the
$RXTE$ spectrum.  It should be noted that this method is based on the
assumption that the cooling in the post-shock region is entirely due to the
bremsstrahlung process and the contribution from cyclotron cooling is not
significant, hence it is more suitable for IPs.  In the case of polars, it
can only provide a lower limit to the mass of the white dwarf.  On the
other hand, the mass determined using the multi-temperature plasma model is
perhaps more reliable, since it takes in to account the cyclotron cooling
effects (see \S 4.1).  From the observed duration of the eclipse in
V1432~Aql, Mukai et al. (2003a) estimated an upper limit of 0.67M$_{\odot}$
for white dwarf mass.  Our mass estimate obtained using the X-ray continuum
and the line spectroscopy methods differ significantly from Mukai's
estimate, and is close to that of 0.98M$_{\odot}$ reported by Ramsay (2000)
using spectral fitting of the X-ray continuum.

\subsection{Blackbody Temperature}

A strong blackbody component dominates the soft X-ray energies below 0.8
keV irrespective of the models used for the higher energy spectra.  This
component has an average temperature between 80--90 eV, which is
significantly higher than what is generally observed in polars (20--40 eV)
(Ramsay et al. 2004a and references therein).  A somewhat higher blackbody
temperature in the range of 50--60 eV has been reported in a few soft IPs,
namely, V405~Aur and PQ~Gem by Haberl \& Motch (1995) using $ROSAT$ data.
Recently, de Martino et al. (2004) have reported the blackbody temperatures
in the range of 60--100 eV for three soft IPs (V405~Aur, PQ~Gem and
V2400~Oph) using broad band (0.1--90 keV) observations with $BeppoSAX$.  A
blackbody temperature of 86 eV is reported for another soft IP 1RXS
J154814.5-452845 by Haberl, Motch \& Zickgraf (2002).  Thus, V1432~Aql
shows a blackbody temperature similar to that observed in soft IPs.  High
blackbody temperature indicates that the accretion is taking place over a
much smaller fractional area on the surface of the white dwarf.  But this
is a too simplistic picture, since accretion regions in polars are
generally smaller than that in IPs.  The blackbody temperature is also high
for soft IPs, but they cannot necessarily have smaller accretion regions
than polars.

\subsection{Energy Balance}

The standard model for MCVs predicts a ratio of intrinsic soft X-ray
luminosity to the hard component to be $\sim$0.5, including the cyclotron
emission component (Lamb \& Masters 1979 and King \& Lasota 1979).  Several
polars show this ratio to be higher than that predicted by the standard
model, leading to the well-known `soft X-ray excess' problem.  On the other
hand, IPs typically show a ratio in agreement with the standard model or
somewhat lower than that predicted.  A study of this ratio can provide
valuable information about the energy balance in these systems.  We have
listed in Table 4 the unabsorbed soft and hard X-ray fluxes and the
corresponding luminosities (assuming d=230 pc; Watson et al. 1995) in the
0.2--1 keV and 1--10 keV energy bands, respectively.  We have also
calculated the unabsorbed bolometric flux and luminosity for the blackbody
component and listed them in Table 4.  Since the blackbody component is
assumed to be optically thick emission, one has to apply a geometric
correction factor $sec(\theta)$ to account for projection effects, where
$\theta$ is the angle between normal to the accretion region and the line
of sight.  For an inclination angle ($i$) of 75$^\circ$ and magnetic
co-latitude ($\beta$) of 15$^\circ$ (Mukai et al. 2003a), the value of
$\theta$ (=$i$-$\beta$) is 60$^\circ$, which corresponds to a correction
factor of 2.  We apply this correction factor for estimating the average
unabsorbed bolometric luminosity due to the blackbody (2.0 $\times$
10$^{31}$ ergs s$^{-1}$).  The unabsorbed hard X-ray flux in the 1--10 keV
energy band is 2.13 $\times$ 10$^{-11}$ ergs cm$^{-2}$ s$^{-1}$ and
corresponds to luminosity of 1.27 $\times$ 10$^{32}$ ergs s$^{-1}$.  This
gives a ratio of the soft to hard X-ray luminosity of $\sim$0.16, a
comparatively lower value than predicted by the standard model. The
photo-ionized plasma model gives a comparatively higher value for the hard
X-ray luminosity and a similar value for the soft X-ray bolometric
luminosity as obtained from multi-temperature plasma model, thus, further
reducing the ratio for soft to hard X-ray luminosity to 0.14.  These values
of ratio for soft to hard X-ray luminosity should be taken as upper limits
since the contribution from cyclotron emission is not included because the
cyclotron spectra were not available.  Ramsay et al. (1994) showed that
this ratio is correlated with magnetic field strength, and a low observed
value for the ratio indicates a low value for magnetic field strength.  A
recent study by Ramsay \& Cropper (2004b) using $XMM$-$Newton$ data for a
sample of polars shows no evidence of such a correlation and instead shows
that $\sim$53\% of all the polars studied in the sample show the ratio to
be $<$1.

\subsection{The Shock Height}

Theoretical models predict that the shock height in an
accretion column can be described in terms of the mass accretion rate
and the fractional accreting area by the relationship
\begin{equation}
H = 5.45 \times 10^{8} \dot{M}_{16}^{-1} f_{-2} M_{WD}^{3/2} R_{WD}^{1/2}
\end{equation}
(Frank, King \& Raine 1992), where $\dot{M}_{16}$ is the mass accretion
rate in units of 10$^{16}$ g s$^{-1}$, $f_{-2}$ is the fractional
area in units of 10$^{-2}$, M$_{WD}$ and R$_{WD}$ are the mass and the
radius of white dwarf in solar units respectively.

The accretion rate $\dot{M}$ can be estimated assuming that the accretion
luminosity is emitted mostly in the X-rays and is given by,
\begin{equation} L_{acc} = \frac{GM_{WD}\dot{M}}{R_{WD}} \end{equation}
This requires an estimation for the accretion luminosity L$_{acc}$ and
R$_{WD}$.  The unabsorbed X-ray luminosity obtained from the $RXTE$
spectral fit in the extended high energy band of 2--60 keV is $1.1
\times 10^{33}$ ergs s$^{-1}$.  The spectral fit to the $XMM$-$Newton$
data using the same model gives an average soft X-ray bolometric
luminosity of $5.6 \times 10^{31}$ ergs s$^{-1}$ after scaling the
normalization with respect to the $RXTE$ spectral fit.  The total
accretion luminosity for the soft and hard X-ray energy bands is $1.1
\times 10^{33}$ ergs s$^{-1}$.  The radius of the white dwarf,
R$_{WD}$, can be estimated using the mass-radius relationship as given
by Nauenberg (1972) and the estimated mass of the white dwarf
(1.2M$_{\odot}$; see \S 4.1).  This gives a corresponding radius of the
white dwarf of 3.8 $\times 10^{8}$ cm ($\sim$0.005R$_{\odot}$).
Substituting these values in equation (2) and solving it for $\dot{M}$
gives a mass accretion rate of $2.7 \times 10^{15}$ g s$^{-1}$.

The above estimated values of the parameters R$_{WD}$, $\dot{m}$, and
$\dot{M}$ enable us to calculate the value of the fractional area f as $1.5
\times 10^{-3}$.  Finally, substituting the values of $\dot{M}$, $f$,
M$_{WD}$ and R$_{WD}$ in equation (1), gives a value for the shock height
as H=$2.9 \times 10^7$ cm (0.08R$_{WD}$).  However, propagating the errors
on the accretion luminosity, the total accretion rate, the mass and radius
of the white dwarf provide an upper limit on shock height to be $\sim$$5
\times 10^7$ cm.  James et al. (2002) estimate an upper bound on shock
height to be $5.4 \times 10^7$ cm for the PQ~Gem.  The higher shock height
in PQ~Gem is due to the larger fractional accreting area ($f$=$9 \times
10^{-3}$), since the mass and radius of the white dwarf are about the same
for both sources.  A smaller accreting area for V1423~Aql is consistent
with a comparatively higher blackbody temperature (88 eV) than that of
PQ~Gem (56 eV; de Martino et al. 2004).

\section{Conclusions}
From our timing study of X-ray and optical emission, and spectral
study of X-ray emission of V1432~Aql, we draw the following
conclusions:
\begin{enumerate}

\item The power spectrum in the 0.1--2 keV energy band obtained from the
combined 1 year long $ROSAT$ observations shows a peak at the period
($\omega$) of 12150.72$\pm$1.14~s that is close to the value known from
optical photometry and identical to the spin period of the system.  A
significant power is also detected in the $\Omega$+$\omega$ component.

\item The power spectrum obtained from the $XMM$-$Newton$ observation
shows a dominant peak at $\sim$4050~s that is likely
due to the complex X-ray light curve with a triple hump profile.  The
strict recurrence of X-ray features at the period of 12150~s suggest
that 12150~s is the likely spin period instead of 4050~s.

\item The power spectrum from the optical circular polarization data
shows prominent peaks corresponding to the first harmonic of the spin
period as well as its one day alias.  The fundamental period and several other
harmonics are also seen.  Two maxima are seen in circular polarization
data in each spin cycle, which explains why the first harmonic of the
spin period is more prevalent than the fundamental.

\item The spectral analysis and observed intensity of the H-like to
the He-like
Fe K$_{\alpha}$ lines indicate a high mass for the white dwarf.

\item Variations seen in the X-ray intensity in the $ROSAT$ data
can be explained using a two pole accretion model in
which two accretion foot prints are not anti-podal.  However, the
triple hump profile seen in the $XMM$-$Newton$ and the $ASCA$ light
curves requires three hot spots on the surface of the white dwarf.
Supportive evidence for the three spots comes from the X-ray phase
resolved spectroscopy and the polarization data.

\item The X-ray spectral study suggests the presence of a strong black
body component with an average temperature of $\sim$88 eV.  This is
significantly higher than commonly observed in polars and is close to
that observed in soft IPs.

\item The originating site for the 6.4 keV Fe fluorescent line in V1432~Aql
most likely contains both neutral (cold) and partially ionized (warm) Fe ions.
Velocity broadening can also be responsible for 6.4 keV line width but
present data can not distinguish between the two possibilities.

\item Both the multi-temperature plasma and the photo-ionization
plasma models fit the spectral data with the photo-ionization model
providing a slightly better fit.  The photo-ionization model, however,
has several parameters that need a higher spectral resolution for
their determination.

\end{enumerate}

\acknowledgments

This research has made use of data obtained from the High Energy
Astrophysics Science Archive Research Center (HEASARC), provided by NASA's
Goddard Space Flight Center.  V.R.R is pleased to acknowledge partial
support from the Kanwal Rekhi Scholarship of the TIFR Endowment Fund.  The
research of P.E.B. was supported by contract number NAG5-12413 to STScI.
The authors wish to thank G. Ramsay and M. Cropper for providing their
program code for the multi-temperature plasma model.  We thank A.
Kinkhabwala for his useful suggestions on spectral fitting of the
photo-ionized plasma model.

\clearpage

\clearpage

\clearpage

\clearpage

\clearpage

\clearpage

\clearpage

\clearpage

\clearpage

\clearpage

\clearpage

\clearpage

\clearpage

\clearpage

\begin{deluxetable}{ccccc}
\tabletypesize{\scriptsize}
\tablecaption{Summary of the X-ray Observations
of V1432~Aquilae.  \label{tbl-1}}
\tablewidth{0pt}
\tablehead{
\colhead{Start Time} & \colhead{End Time} & \colhead{Satellite}   &
\colhead{Exposure} &
\colhead{Mean Count rate\tablenotemark{a,b}}  \\
\colhead{(UT)} & \colhead{(UT)} &    &
\colhead{ks}   & \colhead{(Count s$^{-1}$)}    \\
\colhead{(1)} & \colhead{(2)} & \colhead{(3)} & \colhead{(4)} &
\colhead{(5)}
}
\startdata
1992 Oct 08, 2001 & Oct 30, 1732 & $ROSAT$ & 29.6 & 0.065  \\
1993 Mar 31, 0723 & Apr 02, 0935 & $ROSAT$ & 38.2 & 0.040  \\
1993 Oct 13, 2252 & Oct 18, 1007 & $ROSAT$ & 23.5 & 0.055  \\
1997 Oct 27, 2025 & Oct 29, 0010 & $ASCA$ & 34.2 & 0.04  \\
2001 Oct 09, 0051 & Oct 09, 0755 & $XMM$ & 25.2 & 0.7    \\
2002 Jul 14, 0853 & Jul 14, 1313 & $RXTE$  & 9.4 & 13.0  \\
2002 Jul 15, 1016 & Jul 15, 1432 & $RXTE$ & 10.2 & 10.0   \\
\enddata

\tablenotetext{a}{The 0.1--2 keV energy band for the $ROSAT$, the 2--10
keV energy band for $ASCA$, the 2--20 keV energy band for $RXTE$
and 0.2--10 keV energy band for $XMM$.}
\tablenotetext{b}{RXTE count rates have been normalized for 5 PCUs
(see the text).}

\end{deluxetable}


\begin{deluxetable}{cccccccc}
\tabletypesize{\scriptsize}
\tablecaption{Best fit values of spectral parameters. \label{tbl-2}}
\tablewidth{0pt}
\tablehead{
\colhead{Spin} &
\colhead{$N_{H1}$\tablenotemark{a}}    &
\colhead{$N_{H2}$\tablenotemark{b}}   &
\colhead{Covering} &
\colhead{Abundance\tablenotemark{c}} &
\colhead{Normalization\tablenotemark{d}} & \colhead{A$_{Fe}$\tablenotemark{e}}
& \colhead{$\chi^2_{min}$($\nu$)\tablenotemark{f}}  \\
\colhead{Phase} & \colhead{($\times 10^{21}$ cm$^{-2}$)} &
\colhead{($\times 10^{23}$ cm$^{-2}$)}  &
\colhead{fraction} &   & \colhead{($\times 10^{-7}$)}  &   &
}
\startdata
Average & $1.7^{+0.3}_{-0.3}$ & $1.3^{+0.2}_{-0.1}$ & $0.65^{+0.01}_{-0.01}$ & $0.7^{+0.2}_{-0.2}$ & 11.2 & $4.3^{+0.8}_{-0.8}$ & 1.35(736) \\
0.18--0.30 & $2.6^{+0.5}_{-0.4}$ & $1.4^{+0.3}_{-0.3}$ & $0.64^{+0.02}_{-0.03}$ & $0.8^{+0.2}_{-0.4}$ & 10.2 & $4.5^{+1.8}_{-2.3}$ & 1.06(275)  \\
0.30--0.40 & $0.9^{+0.6}_{-0.7}$ & $1.2^{+0.5}_{-0.5}$ & $0.59^{+0.04}_{-0.07}$ & $1.6^{+1.1}_{-0.8}$ & 4.59 & $3.1^{+2.2}_{-2.1}$ & 1.13(166)  \\
0.40--0.54 & $2.6^{+0.3}_{-0.3}$ & $1.4^{+0.3}_{-0.2}$ & $0.62^{+0.01}_{-0.03}$ & $0.9^{+0.2}_{-0.3}$ & 18.9 & $8.7^{+2.5}_{-2.2}$ & 1.19(507)  \\
0.54--0.73 & $2.0^{+0.5}_{-0.6}$ & $1.3^{+0.2}_{-0.2}$ & $0.83^{+0.01}_{-0.01}$ & $0.6^{+0.3}_{-0.3}$ & 12.3 & $4.6^{+1.7}_{-1.9}$ & 1.19(318)  \\
0.73--0.90 & $0.0^{+0.1}$ & $0.4^{+0.1}_{-0.1}$ & $0.52^{+0.02}_{-0.02}$ & $0.7^{+0.3}_{-0.3}$ & 15.6 & $6.9^{+1.9}_{-1.9}$ & 1.34(543)  \\
0.90--1.18 & $0.0^{+0.1}$ & $7.6^{+3.9}_{-5.3}$ & $0.79^{+0.11}_{-0.17}$ & $0.1^{+0.5}_{-0.1}$ & 6.3 & $1.9^{+1.9}_{-1.5}$ & 1.30(234)  \\
   & $0.0^{+0.1}$ & $2.8^{+4.2}_{-1.6}$ & $0.49^{+0.26}_{-0.13}$ & 0.7 (fixed) & 2.3 & $1.5^{+1.2}_{-0.8}$ & 1.31(235)  \\
\enddata

\tablenotetext{a}{Absorption due to an extra neutral hydrogen column}
\tablenotetext{b}{Absorption due to partial covering of the X-ray
source by neutral hydrogen column}
\tablenotetext{c}{Fe abundance relative to the solar value}
\tablenotetext{d}{Normalization for the continuum}
\tablenotetext{e}{Fe 6.4 keV line flux in units of 10$^{-5}$ photons
cm$^{-2}$ s$^{-1}$}
\tablenotetext{f}{Reduced $\chi^2_{min}$ for $\nu$ degrees of freedom}

\tablecomments{Data for MOS1 and MOS2 were fitted simultaneously using
the multi-temperature plasma model as described by Cropper
et. al. (1999).  The error bars are with 90\% confidence limit for one
parameter. All values are for the multi-temperature plasma model (Cropper et al.
1999) and a black body component with kT$_{bb}$=$88\pm2$ eV for
the average spectrum and frozen at 88 eV for the phase resolved spectra.}

\end{deluxetable}

\clearpage

\begin{deluxetable}{cccccccccc}
\tabletypesize{\scriptsize}
\tablecaption{Best fit spectral parameters. \label{tbl-3}}
\tablewidth{0pt}
\tablehead{
\colhead{Phase} &
\colhead{kT$_{bb}$} & \multicolumn{5}{c}{Column densities for Fe ions (cm$^{-2}$)}  &
\colhead{$\Gamma$} &
\colhead{$\chi^2_{min}$($\nu$)}   \\
  & \colhead{(eV)} & \colhead{Fe I--XVI} & \colhead{Fe XVII--XIX}  &
\colhead{Fe XX--XIV} & \colhead{Fe XXV} & \colhead{Fe XXVI}    &
    &    &
}
\startdata
Average & $80^{+1}_{-2}$ & $8 \times 10^{18}$ & $3 \times 10^{18}$ & $3 \times 10^{17}$ & $2 \times 10^{18}$ & $3 \times 10^{18}$ & $0.64^{+0.01}_{-0.02}$ &  1.28(740)   \\
\enddata

\tablecomments{All values are for the photo-ionized plasma model as described by
Kinkhabwala et al. (2003)}

\end{deluxetable}


\begin{deluxetable}{cccccccc}
\tabletypesize{\scriptsize}
\tablecaption{Soft and hard X-ray flux and luminosity of V1432~Aql.
\label{tbl-4}}
\tablewidth{0pt}
\tablehead{
\colhead{Phase}  & \multicolumn{3}{c}{Flux\tablenotemark{a}} &
\multicolumn{3}{c}{Luminosity\tablenotemark{b}}  &
\colhead{Luminosity\tablenotemark{c}}  \\
   & \colhead{Soft} & \colhead{Soft} & \colhead{Hard} & \colhead{Soft} &
\colhead{Soft} & \colhead{Hard} & \colhead{Ratio}  \\
   & \colhead{(0.2--1 keV)} & \colhead{Bolometric} & \colhead{(1--10 keV)}  &
\colhead{(0.2--1 keV)} & \colhead{Bolometric} & \colhead{(1--10 keV)}  &
\colhead{Soft/Hard}
}
\startdata
Average    & 0.54 & 0.66 & 2.13 & 0.16 & 0.20 & 1.27 & 0.13(0.16)  \\
0.18--0.30 & 0.34 & 1.09 & 1.92 & 0.10 & 0.33 & 1.15 & 0.09(0.28)  \\
0.30--0.40 & 0.17 & 0.50 & 1.06 & 0.05 & 0.15 & 0.63 & 0.08(0.24)  \\
0.40--0.54 & 0.56 & 1.74 & 3.67 & 0.17 & 0.52 & 2.20 & 0.08(0.24)  \\
0.54--0.73 & 0.07 & 0.13 & 1.99 & 0.02 & 0.04 & 1.19 & 0.02(0.03)  \\
0.73--0.90 & 0.39 & 0.73 & 4.05 & 0.12 & 0.22 & 2.42 & 0.05(0.09)  \\
0.90--1.18 & 0.06 & 0.11 & 0.52	& 0.02 & 0.03 & 0.31 & 0.06(0.10)   \\
\enddata

\tablenotetext{a}{Flux in units of 10$^{-11}$ ergs s$^{-1}$ cm$^{-2}$
(absorption through local absorbers only, with $N_H$ due to ISM set to 0)}
\tablenotetext{b}{Luminosity in units of 10$^{32}$ ergs s$^{-1}$}
\tablenotetext{c}{The values inside the parentheses indicate ratio of the
bolometric luminosities}

\tablecomments{The values listed here are based on the
multi-temperature plasma model.  The soft X-ray luminosity is defined
as $L_{soft}$=$\pi$ $F_{soft}$ sec($\theta$) d$^2$ (Ramsay \& Cropper
2004b), where $F_{soft}$ is unabsorbed flux of the soft X-ray
component and $d$ is the distance of the source.  The hard X-ray
luminosity is defined as $L_{hard}$=4$\pi$ $F_{hard}$ d$^2$, where
$F_{hard}$ is unabsorbed flux of the hard X-ray component (d=230 pc).}

\end{deluxetable}
\end{document}